# Classical Mathematical Models for Description and Prediction of Experimental Tumor Growth

Sébastien Benzekry[a,b], Clare Lamont[b], Afshin Beheshti[b], Amanda Tracz[c], John M.L. Ebos[c], Lynn Hlatky[b], Philip Hahnfeldt[b]

[a]Inria Bordeaux Sud-Ouest, Institut de Mathématiques de Bordeaux, Bordeaux, France

[b]Center of Cancer Systems Biology, GRI, Tufts University School of Medicine, Boston, MA, USA

[c]Department of Medicine, Roswell Park Cancer Institute, Elm & Carlton Streets, Buffalo, NY, 142631

**Corresponding author:** S. Benzekry, sebastien.benzekry@inria.fr

# Abstract

Despite internal complexity, tumor growth kinetics follow relatively simple laws that can be expressed as mathematical models. To explore this further, quantitative and discriminant analyses were performed for the purpose of comparing alternative models for their abilities to describe and predict tumor growth. The models were assessed against data from two *in vivo* experimental systems: an ectopic syngeneic tumor (Lewis lung carcinoma) and an orthotopically xenografted human breast carcinoma.

The models included in the study comprised the exponential (with or without free initial volume), exponential-linear, power law, Gompertz, logistic, generalized logistic and von Bertalanffy models, as well as a model with dynamic carrying capacity. For the breast data, the observed linear dynamics were best captured by the Gompertz and exponential-linear models. The latter also exhibited the highest predictive power for this data set, with excellent prediction scores (≥80%) extending out as far as 12 days in the future. For the lung data, the Gompertz and power law models provided the most parsimonious and parametrically identifiable description. In contrast to the breast data, not one of the models was able to achieve a substantial prediction rate (≥70%) beyond the next day lung data point. In this context, adjunction of *a priori* information on the parameter distribution led to considerable improvement of predictions. For instance, forecast success rates went from 14.9% to 62.7% when using the power law model to predict the full future tumor growth curves, using just three data points.

These results not only have important implications for biological theories of tumor growth and the use of mathematical modeling in preclinical anti-cancer drug investigations, but also may assist in defining how mathematical models could serve as potential prognostic tools in the clinic.

# Author Summary

Tumor growth curves display relatively simple time curves that can be quantified using mathematical models. Herein we exploited two experimental animal systems to assess the descriptive and predictive power of nine classical tumor growth models.

Several goodness-of-fit metrics and a dedicated error model were employed to rank the models for their relative descriptive power. We found that the model with the highest descriptive power was not necessarily the most predictive one. The breast growth curves had a linear profile that allowed good predictability. Conversely, not one of the models was able to accurately predict the lung growth curves when using only a few data points. To overcome this issue, we considered a method that uses the parameter population distribution, informed from *a priori* knowledge, to estimate the individual parameter vector of an independent growth curve. This method was found to considerably improve the prediction success rates.

These findings may benefit preclinical cancer research by identifying models most descriptive of fundamental growth characteristics. Clinical perspective is also offered on what can be expected from mathematical modeling in terms of future growth prediction.

# Introduction

Neoplastic growth involves a large number of complex biological processes, including regulation of proliferation and control of the cell cycle, stromal recruitment, angiogenesis and escape from immune surveillance. In combination, these cooperate to produce a macroscopic expansion of the tumor volume, raising the prospect of a possible general law for the global dynamics of neoplasia. Quantitative and qualitative aspects of the temporal development of tumor growth can be studied in a variety of experimental settings, including *in vitro* proliferation assays, three-dimensional *in vitro* spheroids, *in vivo* syngeneic or xenograft implants (injected ectopically or orthotopically), transgenic mouse models or longitudinal studies of clinical images. Each scale has its own advantages and drawbacks, with increasing relevance tending to coincide with decreasing measurement precision. The data used in the current study are from two different *in vivo* systems. The first is a syngeneic Lewis lung carcinoma (LLC) mouse model, exploiting a well-established tumor model adopted by the National Cancer Institute in 1972 [1]. The second is an orthotopic human breast cancer xenografted in severe combined immunodeficient (SCID) mice [2].

Tumor growth kinetics has been an object of biological study for more than 60 years (see e.g. [3] as one of the premiere studies) and has been experimentally investigated extensively (see [4] for a thorough review and [5–8] for more recent work). One of the most common findings for animal [9] and human [10–12] tumors alike is that their relative growth rates decrease with time [13]; or equivalently, that their doubling times increase.

These observations suggest that principles of tumor growth might result from general growth laws, often amenable to expression as ordinary differential equations [14]. The utility of these models can be twofold: 1) testing growth hypotheses or theories by assessing their descriptive power against experimental data and 2) estimating the prior or future course of tumor progression [9,15] either as a personalized prognostic tool in a clinical context [16–20], or in order to determine the efficacy of a therapy in preclinical drug development [21,22].

Cancer modeling offers a wide range of mathematical formalisms that can be classified according to their scale, approach (bottom-up versus top-down) or integration of spatial structure. At the cellular scale, agent-based models [23,24] are well-suited for studies of interacting cells and implications on population-scale development, but computational capabilities often limit such studies to small maximal volumes (on the order of the $mm^3$). The tissue scale is better described by continuous partial differential equations like reaction-diffusion models [19,25] or continuum-mechanics based models [26,27], when spatial characteristics of the tumor are of interest. When focusing on scalar data of longitudinal tumor volume (which is the case here), models based on ordinary differential equations are more adapted. A plethora of such models exist, starting from proliferation of a constant fraction of the tumor volume, an assumption that leads to exponential growth. This model is challenged by the aforementioned observations of non-constant tumor doubling time. Consequently, investigators considered more elaborate models; the most widely accepted of which is the Gompertz model. It has been used in numerous studies involving animal [9,28–31] or human [12,15,30,32] data. Other models include logistic [30,33] or generalized logistic [11,31] formalisms. Inspired by quantitative theories of metabolism and its impact on biological growth, von Bertalanffy [34] derived a growth model based on balance equations of metabolic processes. These considerations were recently developed into a general law of biological growth [35] and brought to the field of tumor growth [36,37]. When the loss term is neglected, the von Bertalanffy model reduces to a power law (see [5,38] for applications of this model to tumor growth). An alternative, purely phenomenological approach led other investigators [39] to simply consider tumor growth as divided into two phases: an initial exponential phase then followed by a linear regimen. Recently, influences of the microenvironment have been incorporated into the modeling, an example being the inclusion of tumor neo-angiogenesis by way of a dynamic carrying capacity [40,41].

Although several studies have been conducted using specific mathematical models for describing tumor growth kinetics, comprehensive work comparing broad ranges of mathematical models for their descriptive power against *in vivo* experimental data is lacking (with the notable exception of [30] and a few studies for *in vitro* tumor spheroids [33,42–44]). Moreover, predictive power is very rarely considered (see [42] for an exception, examining growth of tumor spheroids), despite its clear relevance to clinical utility. The aim of the present study is to provide a rational, quantitative and extensive study of the descriptive and predictive power of a broad class of mathematical models, based on an adapted quantification of the measurement error (uncertainty) in our data. As observed by others [45], specific data sets should be used rather than average curves, and this is the approach we adopted here.

In the following sections, we first describe the experimental procedures that generated the data and define the mathematical models. Then we introduce our methodology to fit the models to the data and assess their descriptive and predictive powers. We conclude by presenting the results of our analysis, consisting of: 1) analysis of the measurement error and derivation of an appropriate error model, subsequently used in the parameters estimation procedure, 2) comparison of the descriptive power of the mathematical models against our two datasets, and 3) determination of the predictive abilities of the most descriptive models, with or without adjunction of *a priori* information in the estimation procedure.

# Materials and Methods

## *Ethics statement*

Animal tumor model studies were performed in strict accordance with the recommendations in the Guide for the Care and Use of Laboratory Animals of the National Institutes of Health. Protocols used were approved by the Institutional Animal Care and Use Committee (IACUC) at Tufts University School of Medicine for studies using murine Lewis lung carcinoma (LLC) cells (Protocol: #P11-324) and at Roswell Park Cancer Institute (RPCI) for studies using human LM2-4$^{LUC+}$ breast carcinoma cells (Protocol: 1227M). Institutions are AAALAC accredited and every effort was made to minimize animal distress.

## *Mice experiments*

### *Cell culture*

Murine Lewis lung carcinoma (LLC) cells, originally derived from a spontaneous tumor in a C57BL/6 mouse [46], were obtained from American Type Culture Collection (Manassas, VA). Human LM2-4$^{LUC+}$ breast carcinoma cells are a metastatic variant originally derived from MDA-MD-231 cells and then transfected with firefly luciferase [47]. All cells were cultured in high glucose DMEM (obtained from Gibco Invitrogen Cell Culture, Carlsbad, CA or Mediatech, Manassas, VA) with 10%FBS (Gibco Invitrogen Cell Culture) and 5% $CO_2$.

### *Tumor Injections*

<u>Subcutaneous mouse syngeneic lung tumor model</u>

C57BL/6 male mice with an average lifespan of 878 days were used [48]. At time of injection mice were 6 to 8 weeks old (Jackson Laboratory, Bar Harbor, Maine). Subcutaneous injections of $10^6$ LLC cells in 0.2 ml phosphate-buffered saline (PBS) were performed on the caudal half of the back in anesthetized mice.

Orthotopic human xenograft breast tumor model
LM2-4[LUC+] cells (1 × $10^6$ cells) were orthotopically implanted into the right inguinal mammary fat pads of 6- to 8-week-old female severe combined immunodeficient (SCID) mice obtained from the Laboratory Animal Resource at RPCI, as previously described [2].

*Tumor measurements*

Tumor size was measured regularly with calipers to a maximum of 1.5 cm³ for the lung data set and 2 cm³ for the breast data set. Largest (L) and smallest (w) diameters were measured subcutaneously using calipers and the formula $V = \frac{\pi}{6} w^2 L$ was then used to compute the volume (ellipsoid). Volumes ranged 14 – 1492 mm³ over time spans from 4 to 22 days for the lung tumor model (two experiments of 10 animals each) and 202 – 1902 mm³ over time spans from 18 to 38 days for the breast tumor data (five experiments conducted utilizing a total of 34 animals). Plots of individual growth curves for both data sets are reported in Figure S1.

## *Mathematical models*

For all the models, the descriptive variable is the total tumor volume, denoted by $V$, as a function of time $t$. It is assumed to be proportional to the total number of cells in the tumor. To reduce the number of degrees of freedom, all the models except the exponential $V_0$ model had a fixed initial volume condition. Although the number of cells that actually remain in the established tumor is probably lower than the number of injected cells (∼60-80%), we considered 1 mm³ ($\simeq 10^6$ cells [49], i.e. the number of injected cells) as a reasonable approximation for $V(t = 0)$.

*Exponential-linear models*

The simplest theory of tumor growth presumes all cells proliferate with constant cell cycle duration $T_C$. This leads to exponential growth, which is also valid in the extended cases where either a constant fraction of the volume is proliferating or the cell cycle length is a random variable with exponential distribution (assuming that the individual cell cycle length distributions are independent and identically distributed). As one modification, initial exponential phase can be assumed to be followed by a linear growth phase [39], giving the following Cauchy problem for the volume rate of change (growth rate):

$$\begin{cases} \frac{dV}{dt} = a_0 V, & t \leq \tau \\ \frac{dV}{dt} = a_1, & t > \tau \\ V(t = 0) = V_0 \end{cases} \quad (1)$$

Here, the coefficient $a_0$ is the fraction of proliferative cells times $\ln 2 / T_C$ where $T_C$ is either the constant cell cycle length or the mean cell cycle length (under the assumption of exponentially distributed cell cycle lengths). The coefficient $a_1$ drives the linear phase. Assuming that the solution of the problem (1) is continuously differentiable uniquely determines the value of $\tau$ as $\tau = \frac{1}{a_0} \log\left(\frac{a_1}{a_0 V_0}\right)$. The coefficient $V_0$ denotes the initial volume.

From this formula, three models were considered: a) initial volume fixed to 1 mm³ and no linear phase ($a_1 = +\infty$), referred to hereafter as *exponential 1*, b) free initial volume and no linear phase, referred to as *exponential $V_0$* and c) equation (1) with fixed initial volume of 1 mm³, referred to as the *exponential-linear model*.

*Logistic and Gompertz models*

A general class of models used for quantification of tumor growth kinetics have a sigmoid shape, i.e. an increasing curve with one inflection point that asymptotically converges to a maximal volume, the carrying capacity, denoted here by $K$. This qualitatively reproduces the experimentally observed growth slowdown [9–12] and is consistent with general patterns of organ and organismal growth. The *logistic model* is defined by a linear decrease of the relative growth rate $\frac{1}{V}\frac{dV}{dt}$ in proportion to the volume:

$$\begin{cases} \frac{dV}{dt} = aV\left(1 - \frac{V}{K}\right) \\ V(t=0) = 1\ mm^3 \end{cases} \quad (2)$$

where $a$ is a coefficient related to proliferation kinetics. This model can be interpreted as mutual competition between the cells (for nutrients or space, for instance), by noticing that under this model the instantaneous probability for a cell to proliferate is proportional to $1 - \frac{V}{K}$. The logistic model has been used, for instance, in [30]. Others (such as [11]) have considered a generalization of the logistic equation, defined by

$$\begin{cases} \frac{dV}{dt} = aV\left(1 - \left(\frac{V}{K}\right)^\nu\right) \\ V(t=0) = 1\ mm^3 \end{cases} \quad (3)$$

that will be referred to as the *generalized logistic model*. Equation (3) has the explicit solution

$$V(t) = \frac{V_0 K}{(V_0^\nu + (K^\nu - V_0^\nu)e^{-a\nu t})^{\frac{1}{\nu}}}$$

which also provides an analytic solution to model (2) when $\nu = 1$. When a different parameterization is employed, replacing $a$ by $\beta = \frac{a}{\nu}$), this model converges when $\nu \to 0$ to the *Gompertz model*, defined by

$$\begin{cases} \frac{dV}{dt} = ae^{-\beta t}V \\ V(t=0) = 1\ mm^3 \end{cases} \quad (4)$$

Coefficient $a$ here is the initial proliferation rate (at $V = 1$ mm³) and $\beta$ is the rate of exponential decay of this proliferation rate. Although first introduced in [50] for a different purpose – the description of human mortality for actuarial applications – the Gompertz model became a widely-accepted representation of growth processes in general [51] and of tumor growth in particular. It was first successfully used in this regard [28] before its applicability was confirmed on large animal data sets [9,29] and for human breast data [32]. The essential characteristic of the Gompertz model is that it exhibits exponential decay of the relative growth rate. An analytic formula can be derived for the solution of (4):

$$V(t) = V_0 e^{\frac{a}{\beta}(1-e^{-\beta t})}$$

where we can see that asymptotically, the volume converges to a carrying capacity given by $K = V_0 e^{\frac{a}{\beta}}$.

A unified model deriving these three sigmoidal models from specific biophysical assumptions about different types of cellular interactions can be found in [52].

*Dynamic carrying capacity*

Taking the next step up in complexity brings us to a model that assumes a dynamic (time-dependent) carrying capacity (CC) [40,41] that can be taken, for example, to represent the tumor vasculature. If one assumes that stimulation of the carrying capacity is proportional to the tumor surface, and neglects angiogenesis inhibition, this model can be formulated in terms of two coupled equations:

$$\begin{cases} \dfrac{dV}{dt} = aV \log\left(\dfrac{K}{V}\right) \\ \dfrac{dK}{dt} = bV^{2/3} \\ V(t=0) = 1 \ mm^3, K(t=0) = K_0 \end{cases} \quad (5)$$

and will be referred to as the *dynamic CC model*. It should be noted that this model was first developed with the intent of modeling the effect of anti-angiogenic therapies on tumor growth and not strictly for describing or predicting the behavior of $V$ alone. However, we integrated it into our analysis in order to investigate and quantify whether consideration of a dynamic carrying capacity could benefit these tasks.

*Von Bertalanffy and power law*

Von Bertalanffy [34], followed later on by others [35], proposed to derive general laws of organic growth from basic energetics principles. Stating that the net growth rate should result from the balance of synthesis and destruction, observing that metabolic rates very often follow the law of allometry (i.e. that they scale with a power of the total size) [34] and assuming that catabolic rates are in proportion to the total volume, he derived the following model for growth of biological processes

$$\frac{dV}{dt} = aV^\gamma - bV \quad (6)$$

Employing our usual assumption that $V(t=0) = 1 \ mm^3$, we will refer to this model as the *von Bertalanffy model* (note that others [14,30] often identify this model as the specific case $\gamma = 2/3$, termed "second type growth"). It has already been successfully applied to describe tumor growth [36,37]. More elaborate considerations linking tumor growth, metabolic rate and vascularization leading to equation (6) can be found in [37]. That work also provides expressions of the coefficients in terms of measurable energetic quantities. Explicit solution of the model is given by

$$V(t) = \left(\frac{a}{b} + \left(V_0^{1-\gamma} - \frac{a}{b}\right)e^{-b(1-\gamma)t}\right)^{\frac{1}{1-\gamma}}$$

From the observation that our data does not exhibit a clear saturation phase, a qualitative feature of equation (6), we also considered another model, derived from (6), by neglecting the loss term, i.e. taking $b = 0$. This model will be termed the *power law model*. Pushing further the reasoning of [34] and arguing that the rate of synthesis of new material, in the context of tumor growth, should be proportional to the number of proliferative cells (under the assumption of a constant cell cycle length), this model suggests that the proliferative tissue is proportional to $V^\gamma$. This could be further interpreted as a possible fractional Hausdorff dimension of the proliferative tissue, when viewed as a metric subspace of the full tumor volume (viewed itself as a three-dimensional subset of the three-dimensional Euclidean space). This dimension would be equal to $3\gamma$ and could be less than 3 when $\gamma < 1$. In this interpretation, the case $\gamma = \frac{2}{3}$ (i.e. dimension equal to 2) could correspond to a proliferative rim limited to the surface of the tumor. This implies that the tumor radius — proportional to $V^{1/3}$ — grows linearly in time. Such linear growth of the tumor radius has been experimentally reported for tumor growth, for instance in the case of gliomas [18]. At the other extreme, a three-dimensional proliferative tissue ($\gamma = 1$) represents proliferative cells uniformly distributed within the tumor and leads to exponential growth. Any power $0 < \gamma < 1$ gives a tumor growth with decreasing growth fraction (and thus decreasing relative growth rate), for which the power law model provides a description in terms of a geometrical feature of the proliferative tissue. This model was first used for murine tumor growth description in [38] and was applied to human data in [5].

## *Fit procedures and goodness of fit criteria*

### *Individual approach*

The main method we used to fit the models is based on individual fits for each animal. The underlying statistical framework is to consider the volume data $y_i^j$ for animal $j$ ($1 \leq j \leq J$) at time $t_i^j$ (with $1 < i < I^j$) as realizations of a random variable $Y_i^j$ being generated by a (deterministic) model $M$ (itself dependent on a parameter vector of length $P$ $(\beta_1^j, \ldots, \beta_P^j)$), as perturbed by random effects, assumed to be Gaussian. In mathematical terms:

$$Y_i^j = M(t_i^j, \beta^j) + \sigma^j E_i^j \varepsilon_i^j \qquad (7)$$

where the $\varepsilon_i^j$ are independent reduced centered Gaussian random variables and $\sigma^j E_i^j$ is the standard deviation of the error. Statistical analysis of the measurement error was performed (see the Results section) and resulted in the following expression

$$E_i^j = \begin{cases} (Y_i^j)^\alpha, & Y_i \geq V_m \\ V_m^\alpha, & Y_i < V_m \end{cases}$$

For a given animal $j$ and parameter set $\beta^j$, the likelihood $L(\beta^j)$ is defined as the probability of observing $\left(y_1^j, \ldots, y_{I^j}^j\right)$ under model $M$, parameter set $\beta^j$ and expression (7), i.e. $L(\beta^j) = \mathbb{P}\left(y_1^j, \ldots, y_{I^j}^j | \beta^j\right)$. Considering that maximizing $L$ is equivalent to minimizing $-\ln L$, it leads to a weighted least squares minimization problem with objective defined by

$$\chi^2(\beta^j) = \sum_{i=1}^{I^j} \left(\frac{y_i^j - M(t_i^j, \beta^j)}{E_i^j}\right)^2 \qquad (8)$$

Minimization was performed using the Matlab [53] function *lsqcurvefit* (trust-region algorithm), except for the generalized logistic model, for which the function *fminsearch* (Nelder-Mead algorithm) was employed (see supporting text S1 for more details on the numerical procedures). The resulting best-fit parameter vector was denoted $\hat{\beta}^j$. Standard errors ($se$) of the maximum likelihood estimator, from which confidence intervals can be derived, were used to quantify the reliability of the parameters estimated. These were computed from an *a posteriori* estimate of $\sigma^2$, denoted $NMSE$, and the weighted jacobian matrix of the model for animal $j$, denoted $J^j$, both defined by

$$NMSE^j = \frac{1}{I^j - P}\chi^2(\hat{\beta}^j), \quad J^j = \left(\frac{1}{E_i^j}\frac{\partial M}{\partial \beta_p}(t_i^j, \hat{\beta}_p^j)\right)_{i,p} \qquad (9)$$

From these expressions, normalized standard errors can be approximated, in the context of nonlinear least squares regression, by [54]

$$Cov^j = NMSE^j \left(J^{j^T} J^j\right)^{-1}, \quad se_p^{j,2} = \left(Cov^j\right)_{p,p}, \quad nse_p = \frac{se_p}{\hat{\beta}_p} \times 100 \qquad (10)$$

Different initializations of the algorithm were systematically tested to establish the practical identifiability of the models (see supporting text S2).

From the obtained $\hat{\beta}^j$, we derived various indicators of the goodness of fit. The Akaike Information Criterion (AIC) [55,56] was used to compare models with different numbers of parameters by penalizing those that use greater numbers of parameters. It is defined, up to an additive constant that does not depend on the model, by

$$AIC^j = I^j \log\left(\frac{\chi^2(\hat{\beta}^j)}{I^j}\right) + 2K, \qquad (11)$$

where $K = P + 1$. Due to the limited number of data for a given individual, we also considered a corrected version of the AIC, termed AICc [55,56]:

$$AICc^j = AIC^j + \frac{2K(K+1)}{I^j - K - 1} \qquad (12)$$

The Root Mean Squared Error (RMSE) is another classical goodness of fit criterion that also penalizes the lack of parameter parsimony in a model:

$$RMSE^j = \sqrt{\frac{1}{I^j - P}\chi^2(\hat{\beta}^j)} \qquad (13)$$

Yet another criterion is the coefficient of determination:

$$R^{2,j} = 1 - \frac{\sum_{i=1}^{I^j}\left(y_i^j - M(t_i^j, \hat{\beta}^j)\right)^2}{\sum_{i=1}^{I^j}\left(y_i^j - \bar{y}^j\right)^2} \qquad (14)$$

where $\bar{y}^j$ is the time average of the data points. This metric quantifies how much of the variability in the data is described by the model $M$ and how much the model is better at describing the data than the mere mean value.

Finally, we considered as an additional criterion of validity of a fit the p-value obtained from the Kolmogorov-Smirnov statistical test for normality of the weighted residuals, these being defined by

$$Res_i^j = \frac{y_i^j - M(t_i^j, \hat{\beta}^j)}{E_i^j} \quad (15)$$

*Population approach (mixed-effect models)*

The procedure we explained above considers all the animals within a group to be independent. On the other hand, the mixed-effect approach [57] consists of pooling all the animals together and estimating a global distribution of the model parameters in the population. More precisely, the individual parameter vectors $(\beta^1, ..., \beta^J)$ are assumed to be realizations of a random variable $\beta$ (here taken to be log-normally distributed). The statistical representation is then formula (7) with $\beta$ instead of $\beta^j$, together with

$$\ln \beta \sim \mathcal{N}(\mu, \omega)$$

Two coefficients (the vector $\mu$ of length $p$ and the $p \times p$ matrix $\omega$) represent the total population, instead of the $J$ parameter sets in the individual approach. Combined with an appropriate description of the error variance (derived from our error model), a population likelihood of all the data pooled together can be defined. Usually, no explicit formula can be computed for its expression, making its maximization a more difficult task. This is implemented in a software called Monolix [58], which maximizes the likelihood using the stochastic approximation expectation maximization (SAEM) algorithm [59]. Consistently with our results on the measurement error (see the Results section), the error model (i.e. the expression of $E$ in (7)) was taken to be proportional to a fixed power $\alpha = 0.84$ of the volume, although with a threshold volume $V_m = 0$ (because Monolix does not permit the setting of a threshold volume). From this estimation process, a population $AICc$, denoted $AICc_{pop}$, was defined, using the same formula as (12) and the $AIC$ returned by Monolix.

## *Model prediction methods*

For a given animal $j$ and model $M$, the general setting considered for prediction was to estimate the model's parameter set using only the first $n$ data points and to use these to predict at a depth $d$, i.e. to predict the value at time $t_n^j + d$, provided that a measurement exists at this day (in which case it will be denoted by $y_{n,d}^j$). The resulting bestfit parameter set was denoted $\hat{\beta}_n^j$.

*Prediction metrics and success score*

Goodness of a prediction was quantified using the normalized error between a model prediction and the data point under consideration, defined by

$$NE_{n,d}^j = \left| \frac{y_{n,d}^j - M(t_n^j + d, \hat{\beta}_n^j)}{\sigma E_{n,d}^j} \right| \quad (16)$$

Prediction of a single time point was considered acceptable when the normalized error was lower than three, corresponding to a model prediction within three standard deviations of the measurement error of the data and generating success results in good agreement with direct visual examinations (see Figures S2 and S3). This allowed us to define a prediction score at the level of the population (denoted by $S_{n,d}$), by the proportion of successful predictions among all animals having measurements both at times $t_n$ and $t_{n+d}$ (whose set will be denoted by $\mathcal{J}_{n,d}$ and total number by $J_{n,d}$T ). This metric is formally defined by

$$S_{n,d} = \frac{\#\{NE_{n,d}^j < 3,\ j \in \mathcal{J}_{n,d}\}}{J_{n,d}} \tag{17}$$

We derived then a global score for each model by averaging over all possible values of $n$ and $d$. When the total number of animals over which the success score is computed is small, this could bias the success score (since for instance only one successfully predicted animal could give a success score of 100% if there is only one animal to predict). To lower this bias, we considered a minimal threshold for $J_{n,d}$, arbitrarily taken to 5 animals. The overall mean success is then defined by

$$\text{Overall mean success} = \underset{n,d \mid J_{n,d} \geq 5}{\text{mean}}\ S_{n,d} \times 100 \tag{18}$$

When assessing prediction over the total future curve, thus involving several time points, we considered the median of the normalized errors:

$$NE_{n,glob}^j = \text{median}\left(NE_{n,d}^j,\ 1 \leq d \leq I^j - n\right) \tag{19}$$

together with its associated prediction score $S_{n,glob}$ and population average $NE_{n,glob}$.

The previous metrics being dependent on our underlying measurement error, we also considered the relative error and its population average, defined by

$$RE_{n,d}^j = \left|\frac{y_{n,d}^j - M(t_n^j + d, \hat{\beta}_n^j)}{y_{n,d}^j}\right|, \quad RE_{n,d} = \frac{1}{J_{n,d}} \sum_{j \in \mathcal{J}_{n,d}} RE_{n,d}^j \tag{20}$$

*A priori information*

For each dataset and model, the total population was randomly and equally divided into two groups. Individual fits for the first group (the "learning" group) were performed using all the available data, generating mean values $(\overline{\beta_1}, \dots, \overline{\beta_P})$ and standard deviations $(\omega_1, \dots, \omega_P)$ of a parameter vector $(\beta_1, \dots, \beta_P)$ within the population. This information was then used when estimating the individual parameter set of a given animal from the second group (the "forecast group"), based only on a subset of its data points, by penalizing the sum of squared residuals in the following way

$$\chi_A^2(\beta^j) = \sum_{i=1}^n \left(Res_i^j\right)^2 + \sum_{p=1}^P \left(\frac{\beta_p^j - \overline{\beta_p}}{\omega_p}\right)^2 \tag{21}$$

with $Res_i^j$ defined by (15). This objective replaced the one defined in (8). The procedure was repeated 100 times (i.e. 100 random assignments of the total population between 10 "learning" animals and 10 "forecast" animals). This number was sufficiently large to have reached convergence in the law of large numbers (no significant difference between 20 and 100 replicates, $p > 0.2$ by Student's t-test). Among these simulation replicates, we only considered as significant the cases where $J_{n,d} \geq 5$. For the lung tumor data set, this did not lead to any exclusion for most of the situations, the only exceptions being for $S_{3,5}$ and $S_{3,6}$ where only 89/100 and 72/100 replicates were eligible, respectively. In contrast, for the breast tumor data and depths 1 to 10, respectively 99, 16, 76, 3, 100, 3, 100, 0, 34 and 77 replicates were eligible. Therefore, results of $S_{3,2}$, $S_{3,4}$, $S_{3,6}$, $S_{3,8}$ and $S_{3,9}$ were considered as non-significant and were not reported.

# Results

## *Measurement error*

The following method was used for analysis of the error made when measuring tumor volume with calipers. One volume per time point per cage was measured twice within a few minutes interval. This gave a total of 133 measurements over a wide range of volumes (20.7 – 1429 mm³). These were analyzed by considering the following statistical representation

$$Y = y_T + \sigma E \varepsilon$$

where $Y$ is a random variable whose realizations are the measured volumes, $y_T$ is the true volume, $\varepsilon$ is a reduced centered Gaussian random variable, and $\sigma E$ is the error standard deviation. The two measures, termed $y_1$ and $y_2$, were, as expected, strongly correlated (Figure 1A, $r = 0.98, p < 0.001$). Statistical analysis rejected variance independent of volume, i.e. constant $E$ ($p = 0.004$, $\chi^2$ test) and a proportional error model ($E = Y$) was found only weakly significant ($p = 0.083$, $\chi^2$ test, see Figure 1B). We therefore introduced a dedicated error model, defined by

$$E = \begin{cases} Y^\alpha, & Y \geq V_m \\ V_m^\alpha, & Y < V_m \end{cases} \quad (22)$$

Two main rationales guided this formulation. First, we argued that error should be larger when volume is larger, a fact that is corroborated by larger error bars for larger volumes on growth data reported in the literature (see Figure 4 in [2] for an example among many others). This was also supported by several publications using a proportional error model when fitting growth data (such as [42,60]). Since here such a description of the error was only weakly significant, we added a power to account for lower-than-proportional uncertainty in large measurements. Second, based on our own practical experience of measuring tumor volumes with calipers, for very small tumors, the measurement error should stop being a decreasing function of the volume because of detectability limits. This motivated the introduction of the threshold $V_m$. After exploration of several values of $V_m$ and $\alpha$, we found $\alpha = 0.84, V_m = 83 \ mm^3$ to be able to accurately describe dispersion of the error in our data ($p = 0.196$, $\chi^2$ test, see Figure 1C). This yielded an empirical value of $\hat{\sigma} = 0.21$.

We did not dispose of double measurements for the breast tumor data and the error analysis was performed using the lung tumor data set only. However, the same error model was applied to the breast tumor data, as both relied upon the same measurement technique.

This result allowed quantification of the measurement error inherent to our data and was an important step in the assessment of each model's descriptive power.

## *Descriptive power*

We tested all the models for their descriptive power and quantified their respective goodness of fit, according to various criteria. Two distinct estimation procedures were employed. The first fitted each animal's growth curve individually (minimization of weighted least squares, with weights defined from the error model of the previous section, see Material and Methods). The second method used a population approach and fitted all the growth curves together. Results are reported in Figure 2 and Tables 1 and 2. Parameter values resulting from the fits are reported in Tables 3 and 4.

Figure 2.A depicts the representative fit of a given animal's growth curve for each data set using the individual approach. From visual examination, the exponential 1 (1), logistic (2) and exponential-linear (1) models did not well explain lung tumor growth and the exponential 1 (1) and logistic (2) models did not satisfactorily fit the breast tumor growth data. The other models seemed able to describe tumor growth in a reasonably accurate fashion.

These results were further confirmed by global quantifications over the total population, such as by residuals analysis (Figure 2.C) and global metrics reported in Tables 1 and 2. When considering goodness-of-fit only, i.e. looking at the minimal least squared errors possibly reached by a model to fit the data (metric $\frac{1}{I}\chi^2$ in Tables 1 and 2), the generalized logistic model (3) exhibited the best results for both data sets (first column in Tables 1 and 2). This indicated a high structural flexibility that allowed this model to adapt to each growth curve and provided accurate fits. On the other hand, the exponential 1 (1) and logistic (2) models clearly exhibited poor fits to the data, a result confirmed by almost all the metrics (with the exception of the $AICc$).

### *Influence of the goodness-of-fit metric*

Being able to closely match the data is not the only relevant criterion to quantify the descriptive power of a model since parameter parsimony of the model should also be taken into account. Other metrics were employed that balanced pure goodness-of-fit and the number of parameters (see Materials and Methods for their definitions). Among them, $AICc$ exhibited the strongest penalization for a large number of parameters. However, this metric was in multiple instances in disagreement with the other metrics dealing with parsimony. For this reason, we also reported the values of $AIC$. These were found globally in accordance with the $RMSE$. The $AICc_{pop}$ gave a weaker importance to the number of parameters, due to the large number of data points in the setting of the population approach, since all the animals were pooled together. For the same reason, values of $AICc_{pop}$ were almost identical to values of $AIC_{pop}$ and only the former were reported. Other structural and numerical differences (for instance, the individual approach used a deterministic optimizer while the population approach was based on a stochastic algorithm) also explained the discrepancies between the two approaches. When comparing the results generated by the two approaches, better individual fits were obtained using the individual approach (see Tables S2). Indeed, the population approach is better designed for settings where the number of data points is too low to individually estimate the parameters, which was not our case.

Taking all these considerations into account, we deemed the $RMSE$ metric to be a good compromise and used this criterion for ranking the models in Tables 1 and 2.

*Descriptive power and identifiability of the models for each data set*

Lung data
Five models (generalized logistic (3), Gompertz (4), power law (6), dynamic CC (5) and von Bertalanffy (6)) were found to have similar $RMSE$ (Table 1), suggesting an identical descriptive power among them. However, having one less parameter, the Gompertz and power law models had smaller $AIC$ (and much smaller $AICc$) and should thus be preferred for parsimonious description of subcutaneous tumor growth of LLC cells. Having an additional degree of freedom translated into poor identifiability of the parameters for the generalized logistic (3), dynamic CC (5) and von Bertalanffy (6) models, as indicated by high standard errors on the parameter estimates (last column of Table 3) and low robustness of these estimates with regard to the initialization of the parameters (see the study of practical identifiability of the models in supplementary text S2 and Table S3). The Gompertz model (4) was also supported by the observation that the median value of $\nu$ estimated by the generalized logistic model was close to zero.

Breast data
Superior fitting power was obtained by the exponential-linear model (1), for all but one of the metrics considered ($R^2$, see Table 2). For all the animals, the fits were in the linear phase of the model indicating linear tumor growth dynamics in the range of volumes observed. The Gompertz (4), generalized logistic (3) and power law (6) models still had high descriptive power, with mean $RMSE$ and $AIC$ similar to the exponential-linear model (Table 2). Again, as a consequence of their larger number of parameters, the dynamic CC (5), von Bertalanffy (6) and generalized logistic (3) models exhibited very large standard errors of the parameter estimates as well as large inter-animal variability (Table 3). Consequently, moderate confidence should be attributed to the specific values of the parameters estimated by the fits, although this did not affect their descriptive power.

As a general result for both data sets, all the models with two parameters were found to be identifiable (Tables 2 and 3). This was confirmed by a study of practical identifiability performed by systematically varying the initial condition of the minimization algorithm (see supporting text S2 and Table S3). For the theories that were able to fit (power law (6) and Gompertz (4) models for the lung tumor data and additionally the exponential-linear model (1) for the breast tumor data), the values of the parameters and their coefficient of variability provided a fairly good characterization of the tumor growth curves dynamics and inter-animal variability. In particular, the power $\gamma$ of the power law model identified in the lung tumor data set seemed to accurately represent the growth of the LLC experimental model (low standard errors and coefficient of variation). Results of inter-animal variability suggested a larger heterogeneity of growth curves in the breast tumor data than in the lung tumor data set, which could be explained by the different growth locations (orthotopic versus ectopic).

Taken together, our results show that, despite the complexity of internal cell populations and tissue organization, at the macroscopic scale tumor growth exhibits relatively simple dynamics that can be captured through mathematical models. Models with three parameters, and more specifically the generalized logistic model (3), were found highly descriptive but not identifiable. For description of subcutaneous *in vivo* tumor growth of LLC cells, the Gompertz (4) and power law (6) models were found to exhibit the best compromise between number of parameters and descriptive power. Orthotopic growth of LM2-4$^{LUC+}$ cells showed a clear linear trend in the range of observed volumes, well captured by the exponential-linear (1), power law (6) and Gompertz (4) models.

## *Forecasting tumor growth. Individual curves*

The two models that were shown unable to describe our data in the previous section, namely the exponential 1 (1) and logistic (3) models, were excluded from further analysis. The remaining ones were assessed for their predictive power. The challenge considered was to predict future growth based on parameter estimation performed on a subset of the data containing only $n$ data points (with $n < I^j$ for a given $j$). We refer to the Materials and Methods section for the definitions of prediction metrics and success scores.

### *Models' predictive power for $n = 5$*

The initial scenario considered the prediction of future growth based on the first five data points.

Lung data
Figure 3 presents a representative example of predictions in this setting for a given animal of the lung tumor data set (mouse 2, see Figure S2.A for specific predictions for each of the animals using the Gompertz model (4)). The success criterion that we defined in the Materials and Methods was found to be in agreement with direct visual examination. According to this metric, the power law (6), dynamic CC (5) and von Bertalanffy (6) models seemed able to accurately predict the global future growth curve while the exponential $V_0$ (1), exponential-linear (1), Gompertz (4) and generalized logistic (3) models, although passing close to the next data point, were less accurate for prediction of the remainder of the data.

Quantifications of the goodness of the prediction on the total population, reported in Table 5 (see the metric $S_{5,glob}$) showed that the prediction success depended on the mouse under consideration. Despite the low predictive power of the curve of Figure 3, the Gompertz model (together with the von Bertalanffy model), had the best global score $S_{5,glob}$, predicting 9/20 mice. A more detailed examination of mice for when the Gompertz model (4) failed (Figure S2.A), indicated that most of time the model interpreted too strongly an initial slowdown. This resulted in large underestimation of future data points (see mice 3, 4, 5 and 13 in Figure S2.A). The same predictive pattern and almost identical predictive curves were observed for the von Bertalanffy (6), dynamic CC (5) and power law (6) models. On the other hand, when using the generalized logistic model (3), some growth curves showed a different predictive pattern (see Figure S2.B). Due to the high flexibility already observed in the descriptive study, this model often saturated early to fit the first five data points, resulting in poor future predictions.

Study of short term predictability, for instance at a depth of two days (score $S_{5,2}$, Table 5), showed that no more than an average relative precision of 19% should be expected, for all predictions taken together.

Breast data
Substantial differences and overall better predictability were found for the same setting ($n = 5$) for the breast tumor data. For instance, the average relative precision at a depth of two days was 13%, using the exponential-linear model. This improved predictability was also expressed by a higher $S_{5,glob}$, although caution should be employed in this comparison since the number of points predicted in $S_{5,glob}$ was lower in the breast tumor setting than in the lung tumor setting (see Figures S2.A and S3). Predictions of all animals using the exponential-linear model were reported in Figure S3 and showed that the linear dynamics exhibited by the breast tumor growth curves could explain this better predictability.

### *Variable number of data points used for prediction and prediction depth*

For evaluation of the global predictive properties of the models, we investigated varying the number $n$ of data points used for estimation of the parameters (respectively $3 \leq n \leq 9$ and $3 \leq n \leq 6$ for the lung and breast data sets) and the prediction depth $d$ (respectively $1 \leq d \leq 7$ and $1 \leq d \leq 12$). Results are reported in Figure 4 and Tables 5 and 6.

In contrast with its high descriptive power (Tables 1 and 2), the generalized logistic model (3) was found to have the lowest overall mean success rate with all $(n, d)$ settings pooled together for both data sets (Tables 5 and 6). In this case, high descriptive power and low predictive ability were linked together. Indeed, the generalized logistic model (3) suffered from its flexibility when put in a predictive perspective. As expressed before for the case $n = 5$, the model fitted very well the initial parts of the curves, but this resulted in premature saturation of the tumor growth and eventually low prediction scores (see Figure S2.B). On the other hand, high success scores were obtained when the model was fed with a lot of data ($n$ large, see Figure 4.A). This result emphasizes that a model's high descriptive abilities might not always translate into high predictive power.

Lung data
According to their predictive patterns in the $(n, d)$ plane, the four models von Bertalanffy (6), Gompertz (4), dynamic CC (5) and power law (6) could be grouped together, and only one of them is presented in Figure 4.A (the von Bertalanffy model (6), see Figure S4.A for the predictive patterns of the other models). Interestingly, what occurred with the generalized logistic model (lower predictive power associated to high flexibility) was not observed for the two other three-parameter models (von Bertalanffy (6) and dynamic CC (5)). This indicates a rigidity of these models similar to the power law (6) and Gompertz (4) models, despite an additional degree of freedom. Taken together, these four models had moderate predictive power, with mean overall prediction scores lower than 45%. The exponential $V_0$ (1) and exponential-linear (1) models were found to have even lower predictive power (Table 5), suggesting that the exponential initial phase of the growth in this data set, might not be predictive of future growth. In most of the situations, the prediction success was found to increase with $n$ and decrease with $d$ (see the von Bertalanffy model (6) in Figure 4.A). Whenever this did not occur (such as in the exponential $V_0$ (1) predictive pattern of Figure 4.A) it was, for most of the cases, due to the fact that the two sets of animals predicted in the two situations were different. In other words, if $S_{n,d} > S_{n',d}$ was observed with $n < n'$, the animals in $\mathcal{J}_{n,d}$ were usually different from the ones in $\mathcal{J}_{n',d}$ (see Materials and Methods, Models prediction methods for the definition of $\mathcal{J}_{n,d}$ and see also Figure S2). This did not imply that the same data points were less accurately predicted with $n$ larger. Surprisingly, this last case was nevertheless observed in some rare settings. For instance, with mouse 19, the generalized logistic model successfully predicted the volumes at days 17 and 18 using four data points (corresponding to $(n, d) = (4,3)$ and $(4,4)$), but failed to do so with five data points (corresponding to $(n, d) = (5,2)$ and $(5,3)$), see Figure S4.

Breast data
For the breast tumor data, predictability was found to be higher than in the lung tumor data, with an excellent overall mean prediction success of the exponential-linear model (1) (83.8%, see Table 6). Consequently, this model ranked 20 percentage points higher than the second best model (dynamic CC (5)). The average score of the exponential-linear model (1) resulted from a wide spread predictability in the $(n, d)$ plane (Figure 4.B), with high success rates even at the far future prediction depth with a small number of data points (for instance, all five of the animals having a data point at $t_3 + 12$ were successfully predicted, $S_{3,12} = 100\%$). While the exponential $V_0$ (1) showed low predictive power, the von Bertalanffy (6), Gompertz (4), power law (6) and dynamic CC (5) models were similarly predictive, having relatively good overall mean success rates (ranging from 58.8% to 63.3%, see Table 6).

As a general result, based on the distribution of relative prediction errors (Figure 4, bottom) all the models had a general trend for underestimation of predictions.

*Tumor growth was more predictable in late phases*

Different tumor growth regimens exist within the same growth curve and, in a clinically relevant setting, diagnosis might occur when the tumor is already large. To explore this further, we tested the predictability of the next day data point (or the second next day when using the breast tumor data, because in this case measurements were performed every two days) in two opposite situations: either using the three first available data points (scores $S_{3,1}$ and $S_{3,2}$ and relative errors $RE_{3,1}$ and $RE_{3,2}$) or using the first three of the last four measurements, as quantified by similar metrics denoted $S_{3,1}^f$, $S_{3,2}^f$, $RE_{3,1}^f$ and $RE_{3,2}^f$. Volume ranges predicted were $303 \pm 128$ mm$^3$ and $909 \pm 273$ mm$^3$ in the early phase for the lung tumor and breast tumor data respectively, versus $1245 \pm 254$ mm$^3$ and $1383 \pm 211$ mm$^3$ in the late phases. In this last setting, in order not to artificially inject the information of the first volume being 1 mm$^3$ at day 0, we modified the von Bertalanffy, dynamic CC, generalized logistic, Gompertz, and power law models by fixing their initial times and volumes to the previous measurement (the fifth from the end). Interestingly, the results obtained were substantially different between the two growth phases. Better predictions were obtained when predicting the end of the curve, reaching excellent scores of 12-15/16 animals successfully predicted in the case of the LLC data (and average relative errors smaller or equal than 15%, see Table 5). Similar improvements were observed for the breast tumor data, with a 63% increase from $S_{3,2}$ to $S_{3,2}^f$ for the power law model and up to an 87% increase for the exponential $V_0$ (see the bracketed numbers in Tables 5 and 6). Hence, the late phase of tumor growth appeared more predictable, possibly because of smaller curvatures of the growth curves that led to better identifiability of the models when using a limited number of data points for estimation of the parameters.

Overall, our results showed equivalent predictive power of the von Bertalanffy (6), dynamic CC (5), power law (6), and Gompertz (4) models for prediction of future tumor growth curves of subcutaneous LLC cells, with substantial prediction rates ($\geq$70%) requiring at least four data points and at a depth no larger than one day. The exponential-linear model was better suited for the orthotopic xenograft breast tumor data, with success rates larger than 70% in most of the $(n, d)$ cases, including excellent scores at greater depths.

## *Forecasting tumor growth. A priori information.*

When relatively fewer data points were used, for example with only three, individual predictions based on individual fits were shown to be globally limited for the lung tumor data, especially over a large time frame (Figure 4.A, Table 5). However, this situation is likely to be clinically relevant since few clinical examinations are performed before the beginning of therapy. On the other hand, large databases might be available from previous examinations of other patients and this information could be useful to predict future tumor growth in a particular patient. In a preclinical setting of drug investigation, tumor growth curves of animals from a control group could be available and usable when inferring information on the individual time course of one particular treated animal.

An interesting statistical method that could potentiate this *a priori* information consists in learning the population distribution of the model parameters from a given database and to combine it with the individual parameter estimation from the available restricted data points on a given animal. We investigated this method in order to determine if it could improve the predictive performances of the models. Each dataset was randomly divided into two groups. One was used to learn the parameter distribution (based on the full time curves), while the other was dedicated to predictions (limited number of data points). For a given animal of this last group, no information from his growth curve was used to estimate the *a priori* distributions. The full procedure was replicated 100 times to ensure statistical significance, resulting in respectively 2000 and 3400 fits performed for each model. We refer to the Materials and Methods for more technical details. Results are reported in Figure 5.

Lung data
Predictions obtained using this technique were significantly improved, going from an average success score of 14.9% ± 8.35% to 62.7% ± 11.9% (means ± standard deviations) for prediction of the total future curve with the power law model (6) (see Figure 5.A). This model allowed predictions to be made at large future depths. For instance, predictions 7 days in the future could reasonably be considered (average success rate of 50.6%, power law model (6), see Figure 5.C), while their success rate was very low with direct individual prediction (6.07%). Prediction successes reached 90% (power law model (6)) at the closer horizon of the next day data point ($S_{3,1}$), while success rate was only 57.1% using an individual approach (Figure 5.B). Other small horizon depths also reached excellent prediction scores (Figure 5.C). The largest improvement of success rates for the power law model was observed for $S_{3,3}$ that went from an average score of 6.86% (with standard deviation 7.47) to an average score of 75.2% (with standard deviation 12.9), representing more than an 11-fold increase. We report in Figure S5 the details of predictions with and without *a priori* information for all the animals within a given forecast group from the lung tumor data set (power law model (6)). It can be appreciated how additional information on the parameter distribution in the estimation procedure significantly improved global prediction of the tumor growth curves. The impact of the addition of the *a priori* information was however less important when using more data points for the estimation (results not shown).

Breast data
Due to its already high prediction score without adjunction of *a priori* information, the exponential-linear model did not benefit from the method. For the next day data point of the breast tumor growth curves, predictability was already almost maximal without adjunction of *a priori* information and thus no important impact was observed.

For both data sets, not all the models equally benefited from the addition of *a priori* information (Figure 5). Models having the lowest parameter inter-animal variability, such as the power law (6), Gompertz (4), exponential-linear (1), and exponential $V_0$ (1) models (Table 3), which also had better practical identifiability (Tables 2 and S3), exhibited great benefit. In contrast, the models with three parameters showed only modest benefit or even decrease of their success rates (see $S_{3,glob}$ and $S_{3,1}$ for the von Bertalanffy model (6) on the breast tumor data in Figure 5.B), with the exception of the generalized logistic model (3) on the breast tumor data. In these cases, adjunction of *a priori* information translated into poor enhancement of predictive power because the mean population parameters did not properly capture the average behavior within the population and were therefore not very informative. On the other hand, models such as the power law model (6) on the lung tumor data set, whose coefficient $\gamma$ characterized particularly well the growth pattern (Table 3), had a more informative *a priori* distribution that translated into the highest improvement of predictive power. For the generalized logistic model (3) on the breast data, the mean parameters were able to inform the linear regimen of the growth phase and thus protected the model from too early saturation.

These results demonstrated that addition of *a priori* information in the fit procedure considerably improved the forecast performances of the models, in particular when using a small number of data points and low-parameterized models for data with low predictability, such as the power law model for the lung tumor data set.

## Discussion

### *Error model*

In our analysis, constant variance of the error was clearly rejected and although a proportional error (used by others [33]) was not strictly rejected by statistical analysis ($p = 0.08$), a more adequate error model to our data was developed. However, using a proportional or even constant error model did not significantly affect conclusions as to the descriptive power of the models, identifying the same models (Tables 1, 2) as most adequate for description of tumor growth (results not shown). Nevertheless, the use of an appropriate error model could have important implications in the quantitative assessment of a model's descriptive performance and rejection of inaccurate tumor growth theories. Interestingly, using the same human tumor growth data, Bajzer et al. [60] found the assumption of proportional error variance to favor the Gompertz model for descriptive ability, whereas Vaidya and Alexandro [30] had observed the logistic model to be favored, under a constant-variance assumption. The error model used might additionally have important implications on predictions. Although detailed analysis of the impact of the error model on prediction power is beyond the scope of the present study, we performed a prospective study of predictive properties when using a constant error model on the lung tumor data and found changes in the ranking of the models (results not shown).

### *Theories of growth*

As expected, our results confirmed previous observations [9–11,13,29,32] that tumor growth is not continuously exponential (constant doubling time) in the range of the tumor volumes studied, ruling out the prospect of a constant proliferating fraction. A less expected finding was that the logistic model (linear decay in volume of the relative growth rate) was also unable to describe our data, although similar results have been observed in other experimental systems [31,33]. On the other hand, the Gompertz and power law models could give an accurate and identifiable description of the growth slowdown, for both data sets. More elaborate models such as the generalized logistic, von Bertalanffy, and dynamic CC models could describe them as well. However, their parameters were found not to be identifiable from only tumor growth curves, in the ranges of the observed volumes. Additional data could improve identifiability, such as relating to later growth and saturation details. It should be noted in this case that the dynamic CC model was not designed with the intent to quantify tumor growth, but rather to describe the effects of anti-angiogenic agents on global tumor dynamics. Because the model carries angiogenic parameters that are not directly measureable, or even inferable, from the experimental systems we used, it stands to reason that they would not be easily identifiable from the data. Kinetics under the influence of antiangiogenic therapy might thus provide useful additional information that could render this model identifiable. For the breast tumor experimental system, the slowdown was characterized by linear dynamics and was most accurately fitted by the exponential-linear model. Observed was exponential growth from the number of injected cells (during the unobserved phase) that switched smoothly to a linear phase (exponential-linear model). It should be noted that in the breast tumor data set, no data were available during the initiation phase (below 200 $mm^3$) and only the linear part of a putative exponential-linear growth was observed. Explorations of the kinetics of growth during the initial phase (at volumes below the $mm^3$) are needed for further clarification.

Despite structural similarities, important differences were noted in the parameter estimates between the two experimental models, in agreement with other studies emphasizing differences between ectopic and orthotopic growth [61,62]. Our results and methodology may help to identify the impact on kinetics of the site of implantation, although explicit comparisons could not be made here due to the differences in the cell lines used.

The Gompertz model (exponential decay in time of the relative growth rate) was able to fit both data sets accurately, consistent with the literature [13,15,29,31,33]. One of the main criticisms of the Gompertz model is that the relative tumor growth rate becomes arbitrarily large (or equivalently, the tumor doubling time gets arbitrarily small) for small tumor volumes. Without invoking a threshold this becomes biologically unrealistic. This consideration led investigators [13,63] to introduce the Gomp-exp model that consists in an initial exponential phase followed by Gompertzian growth when the associated doubling time becomes realistic. This approach could also be applied to any decreasing relative growth rate model. We did not consider it in our analysis due to the already large initial volume and the lack of data on the initiation phase where the issue is most relevant.

The power law model was also able to describe the experimental data and appeared as a simple, robust, descriptive and predictive mathematical model for murine tumor growth kinetics. It suggests a general law of macroscopic *in vivo* tumor growth (in the range of the volumes observed): only a subset of the tumor cells proliferate and this subset is characterized by a constant, possibly fractional, Hausdorff dimension. In our results, this dimension (equal to $3\gamma$) was found to be significantly different from two or three ($p < 0.05$ by Student's t-test) in 14/20 mice for the lung tumor data set and 13/34 mice for the breast tumor data, suggesting effectively a fractional dimension. A possible explanation of this feature could come from the fractal nature of the tumor vasculature [64,65], an argument supported by others who have investigated the link between tumor dynamics and vascular architecture [37]. More precisely, the branching nature of the vascularization generates a fractal organization [37,64,65] that could in turn produce a contact surface of fractional Hausdorff dimension. Considering further that the fraction of proliferative cells is proportional to this contact surface (for instance because proliferative cells are limited to an area at fixed distance from a blood vessel or capillary, due to diffusion limitations), this could make the connection between fractality of the vasculature and proliferative tissue. These considerations could therefore provide a mechanistic explanation for the growth rate decay that naturally happens when the dimension of the proliferative tissue is lower than three. Our results were obtained using two particular experimental systems: an ectopic mouse syngeneic lung tumor and an orthotopic human xenograft breast tumor model. Although consistent with other studies that found the power law model adequate for growth of a murine mammary cell line [38] or for description of human mammography density distribution data [5], these remain to be confirmed by human data. This model should also be taken with caution when dealing with very small volumes (at the scale of several cells for instance) for which the relative growth rate becomes very large. Indeed, the interpretation of a fractional dimension then fails, since the tumor tissue can no longer be considered a continuous medium. In this instance, it may be more appropriate to consider exponential growth in this phase [37].

## *Prediction*

Our results showed that a highly descriptive model (associated to large flexibility) such as the generalized logistic model, might not be useful for predictions, while well-adapted rigidity – as provided by the exponential-linear model on the breast tumor data – could lead to very good predictive power. Interestingly, our study revealed that models having low identifiability (von Bertalanffy and dynamic CC) could nevertheless exhibit good predictive power. Indeed, over a limited time span, different parameter sets for a given model could generate the same growth curves, which would be equally predictive.

For the Gompertz model, predictive power might be improved by using possible correlations between the two parameters of this model, as reported by others [15,63,66–68] and suggested by our own parameter estimates ($R = 0.99$ for both data sets, results not shown).

If a backward prediction is desired (for instance for the identification of the inception time of the tumor), the use of exponential growth might be more adapted for the initial, latency phase, e.g. by employment of the Gomp-exp model [13,63].

### *Clinical and preclinical implications*

Translating our results to the clinical setting raises the possibility of forecasting solid tumor growth using simple macroscopic models. Use of *a priori* information could then be a powerful method and one might think of the population distribution of parameters being learned from existing databases of previous patient examinations. However, the very strong improvement of prediction success rates that we obtained partly comes from the important homogeneity of our growth data (in particular the LLC data) that generated a narrow and very informative distribution of some parameters (for instance parameter $\gamma$ of the power law model), which in turn powerfully assisted the fitting procedure. In more practical situations such as with patient data, more heterogeneity of the growth data should be expected that could alter the benefit of the method. For instance, in some situations, growth could stop for arbitrarily long periods of time. These dormancy phases challenge the universal applicability of a generic growth law such as the Gompertz or power law [69]. Description of such dormancy phenomena could be integrated using stochastic models that would elaborate on the deterministic models reviewed here, as was done by others [70] to describe breast cancer growth data using the Gompertz model. Moreover, further information than just tumor volume could be extracted from (functional) imaging devices, feeding more complex mathematical models that could help design more accurate *in silico* prediction tools [18,71].

Our analysis also has implications for the use of mathematical models as valuable tools for helping preclinical anti-cancer research. Such models might be used, for instance, to specifically ascertain drug efficacy in a given animal, by estimating how importantly the treated tumor deviates from its natural course, based on *a priori* information learned from a control group. Another application can be for rational design of dose and scheduling of anti-cancerous drugs [22,72,73]. Although integration to therapy remains to be added (and validated) to models such as the power law, more classical models (exponential-linear [39] or dynamic CC [41]) have begun to predict cytotoxic or anti-angiogenic effects of drugs on tumor growth. Our methods have allowed precise quantification of their respective descriptive and predictive powers, which, in combination with the models' intrinsic biological foundations, could be of value when deciding among such models which best captures the observed growth behaviors in relevant preclinical settings.

## Acknowledgments

We thank Etienne Baratchart from inria in Bordeaux for valuable suggestions and comments and we appreciate the editorial help of Melissa Klumpar from GeneSys Research Institute in Boston. We are also grateful to three anonymous reviewers for their careful examination of the manuscript, comments and suggestions that led to substantial improvement.

# References

1. Talmadge JE, Singh RK, Fidler IJ, Raz A (2007) Murine models to evaluate novel and conventional therapeutic strategies for cancer. Am J Pathol 170: 793–804. doi:10.2353/ajpath.2007.060929.

2. Ebos JML, Lee CR, Cruz-Munoz W, Bjarnason GA, Christensen JG, et al. (2009) Accelerated metastasis after short-term treatment with a potent inhibitor of tumor angiogenesis. Cancer Cell 15: 232–239. doi:10.1016/j.ccr.2009.01.021.

3. Collins VP, Loeffler RK, Tivey H (1956) Observations on growth rates of human tumors. Am J Roentgenol Radium Ther Nucl Med 76: Am J Roentgenol Radium Ther Nucl Med.

4. Steel GG (1977) Growth kinetics of tumours. Clarendon Press. Oxford.

5. Hart D, Shochat E, Agur Z (1998) The growth law of primary breast cancer as inferred from mammography screening trials data. Br J Cancer 78: 382–387.

6. Friberg S, Mattson S (1997) On the growth rates of human malignant tumors: implications for medical decision making. J Surg Oncol: 284–297.

7. Spratt J a S, Meyer JS (1996) Rates of growth of human neoplasms : part II. J Surg Oncol 61: 143–150.

8. Heuser L, Spratt JS, Polk HC (1979) Growth rates of primary breast cancers. Cancer 43: 1888–1894.

9. Laird AK (1965) Dynamics of tumour growth: comparison of growth rates and extrapolation of growth curve to one cell. Br J Cancer 19: 278–291.

10. Steel GG, Lamerton LF (1966) The growth rate of human tumours. Br J Cancer 20: 74–86.

11. Spratt JA, von Fournier D, Spratt JS, Weber EE (1993) Decelerating growth and human breast cancer. Cancer 71: 2013–2019.

12. Akanuma A (1978) Parameter analysis of Gompertzian function growth model in clinical tumors. Eur J Cancer 14: 681–688.

13. Wheldon TE (1988) Mathematical models in cancer research. Hilger Bristol.

14. Gerlee P (2013) The model muddle: in search of tumor growth laws. Cancer Res 73: 2407–2411. doi:10.1158/0008-5472.CAN-12-4355.

15. Norton L, Simon R, Brereton HD, Bogden AE (1976) Predicting the course of Gompertzian growth. Nature 264: 542–544.

16. Colin T, Iollo A, Lombardi D, Saut O (2010) Prediction of the evolution of thyroidal lung nodules using a mathematical model. ERCIM News: 37–38.

17. Ribba B, Kaloshi G, Peyre M, Ricard D, Calvez V, et al. (2012) A tumor growth inhibition model for low-grade glioma treated with chemotherapy or radiotherapy. Clin Cancer Res 18: 5071–5080. doi:10.1158/1078-0432.CCR-12-0084.

18. Baldock AL, Rockne RC, Boone AD, Neal ML, Hawkins-Daarud A, et al. (2013) From patient-specific mathematical neuro-oncology to precision medicine. Front Oncol 3: 62. doi:10.3389/fonc.2013.00062.

19. Wang CH, Rockhill JK, Mrugala M, Peacock DL, Lai A, et al. (2009) Prognostic significance of growth kinetics in newly diagnosed glioblastomas revealed by combining serial imaging with a novel biomathematical model. Cancer Res 69: 9133–9140. doi:10.1158/0008-5472.CAN-08-3863.

20. Portz T, Kuang Y, Nagy JD (2012) A clinical data validated mathematical model of prostate cancer growth under intermittent androgen suppression therapy. AIP Adv 2: 011002. doi:10.1063/1.3697848.

21. Bernard A, Kimko H, Mital D, Poggesi I (2012) Mathematical modeling of tumor growth and tumor growth inhibition in oncology drug development. Expert Opin Drug Metab Toxicol 8: 1057–1069. doi:10.1517/17425255.2012.693480.

22. Simeoni M, De Nicolao G, Magni P, Rocchetti M, Poggesi I (2013) Modeling of human tumor xenografts and dose rationale in oncology. Drug Discov Today Technol 10: e365–72. doi:10.1016/j.ddtec.2012.07.004.

23. Drasdo D, Höhme S (2005) A single-cell-based model of tumor growth in vitro: monolayers and spheroids. Phys Biol 2: 133–147.

24. Gao X, McDonald JT, Hlatky L, Enderling H (2013) Acute and fractionated irradiation differentially modulate glioma stem cell division kinetics. Cancer Res 73: 1481–1490. doi:10.1158/0008-5472.CAN-12-3429.

25. Gatenby RA, Gawlinski ET (1996) A reaction-diffusion model of cancer invasion. Cancer Res 56: 5745–5753.

26. Ambrosi D, Mollica F (2003) Mechanical models in tumour growth. In: Preziosi L, editor. Cancer Modelling and Simulation. CRC Press. pp. 142–166.

27. Bresch D, Colin T, Grenier E, Ribba B, Saut O (2010) Computational Modeling of Solid Tumor Growth: The Avascular Stage. SIAM J Sci Comput 32: 2321. doi:10.1137/070708895.

28. Casey AE (1934) The experimental alteration of malignancy with an homologous mammalian tumor material : I. Results with intratesticular inoculation. Am J Cancer 21: 760–775.

29. Laird AK (1964) Dynamics of tumour growth. Br J Cancer 13: 490–502.

30. Vaidya VG, Alexandro FJ (1982) Evaluation of some mathematical models for tumor growth. Int J Biomed Comput 13: 19–36.

31. Michelson S, Glicksman a S, Leith JT (1987) Growth in solid heterogeneous human colon adenocarcinomas: comparison of simple logistical models. Cell Tissue Kinet 20: 343–355.

32. Norton L (1988) A Gompertzian model of human breast cancer growth. Cancer Res 48: 7067–7071.

33. Marusić M, Bajzer Z, Freyer JP, Vuk-Pavlović S (1994) Analysis of growth of multicellular tumour spheroids by mathematical models. Cell Prolif 27: 73–94.

34. Bertalanffy L Von (1957) Quantitative laws in metabolism and growth. Q Rev Biol 32: 217–231.

35. West GB, Brown JH, Enquist BJ (2001) A general model for ontogenetic growth. Nature 413: 628–631. doi:10.1038/35098076.

36. Guiot C, Degiorgis PG, Delsanto PP, Gabriele P, Deisboeck TS (2003) Does tumor growth follow a "universal law"? J Theor Biol 225: 147–151. doi:10.1016/S0022-5193(03)00221-2.

37. Herman AB, Savage VM, West GB (2011) A quantitative theory of solid tumor growth, metabolic rate and vascularization. PLoS One 6: e22973. doi:10.1371/journal.pone.0022973.

38. Dethlefsen LA, Prewitt JM, Mendelsohn ML (1968) Analysis of tumor growth curves. J Natl Cancer Inst 40: 389–405.

39. Simeoni M, Magni P, Cammia C, De Nicolao G, Croci V, et al. (2004) Predictive pharmacokinetic-pharmacodynamic modeling of tumor growth kinetics in xenograft models after administration of anticancer agents. Cancer Res 64: 1094–1101. doi:10.1158/0008-5472.CAN-03-2524.

40. Wilson S, Grenier E, Wei M, Calvez V, You B, et al. (2013) Modeling the synergism between the anti-angiogenic drug sunitinib and irinotecan in xenografted mice. PAGE 22. p. 2826.

41. Ribba B, Watkin E, Tod M, Girard P, Grenier E, et al. (2011) A model of vascular tumour growth in mice combining longitudinal tumour size data with histological biomarkers. Eur J Cancer 47: 479–490. doi:10.1016/j.ejca.2010.10.003.

42. Marušić M, Vuk-Pavlović S (1993) Prediction power of mathematical models for tumor growth. J Biol Syst 1: 69–78.

43. Olea N, Villalobos M, Nuñez MI, Elvira J, Ruiz de Almodóvar JM, et al. (1994) Evaluation of the growth rate of MCF-7 breast cancer multicellular spheroids using three mathematical models. Cell Prolif 27: 213–223.

44. Wallace DI, Guo X (2013) Properties of tumor spheroid growth exhibited by simple mathematical models. Front Oncol 3: 51. doi:10.3389/fonc.2013.00051.

45. Retsky MW, Swartzendruber DE, Wardwell RH, Bame PD (1990) Is Gompertzian or exponential kinetics a valid description of individual human cancer growth? Med Hypotheses 33: 95–106.

46. Bertram JS, Janik P (1980) Establishment of a cloned line of Lewis Lung Carcinoma cells adapted to cell culture. Cancer Lett 11: 63–73.

47. Ebos JML, Lee CR, Bogdanovic E, Alami J, Van Slyke P, et al. (2008) Vascular endothelial growth factor-mediated decrease in plasma soluble vascular endothelial growth factor receptor-2 levels as a surrogate biomarker for tumor growth. Cancer Res 68: 521–529. doi:10.1158/0008-5472.CAN-07-3217.

48. Kunstyr I, Leuenberger HG (1975) Gerontological data of C57BL/6J mice. I. Sex differences in survival curves. J Gerontol 30: 157–162.

49. Spratt JS, Meyer JS, Spratt JA (1995) Rates of growth of human solid neoplasms: Part I. J Surg Oncol 60: 137–146.

50. Gompertz B (1825) On the nature of the function expressive of the law of human mortality, and on a new mode of determining the value of life contingencies. Phil Trans R Soc B 115: 513–583. doi:10.1098/rstl.1825.0026.

51. Winsor C (1932) The Gompertz curve as a growth curve. Proc Natl Acad Sci U S A 18: 1–8.

52. Mombach J, Lemke N, Bodmann B, Idiart M (2002) A mean-field theory of cellular growth. Eur Lett 59: 923–928.

53. The Mathworks Inc. (2013) Matlab with statistics and optimization toolboxes.

54. Seber GA, Wild CJ (2003) Nonlinear regression. Wiley-Interscience.

55. Burnham KP, Anderson DR (2002) Model selection and multimodel inference: a practical information-theoretic approach. Springer.

56. Motulsky H, Christopoulos A (2004) Fitting models to biological data using linear and nonlinear regression. Oxford University Press.

57. Ribba B, Holford NH, Magni P, Trocóniz I, Gueorguieva I, et al. (2014) A review of mixed-effects models of tumor growth and effects of anticancer drug treatment used in population analysis. CPT Pharmacometrics Syst Pharmacol 3: e113. doi:10.1038/psp.2014.12.

58. Lixoft (2013) Monolix.

59. Kuhn E, Lavielle M (2005) Maximum likelihood estimation in nonlinear mixed effects models. Comp Stat Data An 49: 1020–1038. doi:10.1016/j.csda.2004.07.002.

60. Bajzer Ž, Vuk-Pavlović S, Huzak M (1997) Mathematical modeling of tumor growth kinetics. In: Adam JA, Bellomo N, editors. A Survey of Models for Tumor-Immune System Dynamics. Birkhäuser Boston. pp. 89–133. doi:10.1007/978-0-8176-8119-7_3.

61. Tsuzuki Y, Carreira CM, Bockhorn M, Xu L, Jain RK, et al. (2001) Pancreas microenvironment promotes VEGF expression and tumor growth: novel window models for pancreatic tumor angiogenesis and microcirculation. Lab Invest 81: 1439–1451. doi:10.1038/labinvest.3780357.

62. Ahn KS, Jung YS, Kim J, Lee H, Yoon SS (2001) Behavior of murine renal carcinoma cells grown in ectopic or orthotopic sites in syngeneic mice. Tumour Biol 22: 146–153. doi:50609.

63. Demicheli R, Foroni R, Ingrosso A (1989) An exponential-Gompertzian description of LoVo cell tumor growth from in vivo and in vitro data. Cancer Res 49: 6543–6546.

64. Gazit Y, Baish JW, Safabakhsh N, Leunig M, Baxter LT, et al. (1997) Fractal characteristics of tumor vascular architecture during tumor growth and regression. Microcirculation 4: 395–402.

65. Baish JW, Jain RK (2000) Fractals and cancer. Cancer Res 60: 3683–3688.

66. Brunton GF, Wheldon TE (1977) Prediction of the complete growth pattern of human multiple myeloma from restricted initial measurements. Cell Tissue Kinet 10: 591–594.

67. Brunton G, Wheldon T (1978) Characteristic species dependent growth patterns of mammalian neoplasms. Cell Tissue Kinet 11: 161–175.

68. Brunton G, Wheldon T (1980) The Gompertz equation and the construction of tumour growth curves. Cell Prolif: 455–460.

69. Retsky M (2004) Universal law of tumor growth. J Theor Biol 229: 289. doi:10.1016/j.jtbi.2004.04.008.

70. Speer JF, Petrosky VE, Retsky MW, Wardwell RH (1984) A stochastic numerical model of breast cancer growth that simulates clinical data. Cancer Res 44: 4124–4130.

71. Cornelis F, Saut O, Cumsille P, Lombardi D, Iollo A, et al. (2013) In vivo mathematical modeling of tumor growth from imaging data: Soon to come in the future? Diagn Interv Imaging 94: 593–600. doi:10.1016/j.diii.2013.03.001.

72. Swierniak A, Kimmel M, Smieja J (2009) Mathematical modeling as a tool for planning anticancer therapy. Eur J Pharmacol 625: 108–121. doi:10.1016/j.ejphar.2009.08.041.

73. Barbolosi D, Freyer G, Ciccolini J, Iliadis A (2003) Optimisation de la posologie et des modalités d'administration des agents cytotoxiques à l'aide d'un modèle mathématique. B Cancer 90: 167–175.

Table 1: Fit performances of the tumor growth models for the lung data

| Model | $\frac{1}{I}\chi^2$ | AIC | AICc | AICc$_{pop}$ | RMSE | R$^2$ | p > 0.05 | # |
|---|---|---|---|---|---|---|---|---|
| Generalized logistic | **0.12 (0.019 - 0.42) [1]** | -13 (-30 - 0.98) [3] | 0.7 [6] | 2114 [6] | **0.4 (0.17 - 0.82) [1]** | **0.98 (0.94 - 1)** | 100 | 3 |
| Gompertz | 0.155 (0.019 - 0.67) [4] | **-13.4 (-32 - 2.4) [1]** | **-7.62 [1]** | 2108 [5] | 0.41 (0.16 - 0.93) [2] | 0.97 (0.82 - 1) | 100 | 2 |
| Power law | 0.155 (0.016 - 0.71) [5] | -13.4 (-34 - 2.9) [2] | -7.59 [2] | 2091 [2] | 0.41 (0.15 - 0.96) [3] | 0.96 (0.78 - 1) | 100 | 2 |
| Dynamic CC | 0.136 (0.013 - 0.61) [2] | -12.5 (-32 - 3.5) [4] | 1.19 [7] | **2063 [1]** | 0.42 (0.14 - 0.96) [4] | 0.97 (0.82 - 1) | 100 | 3 |
| Von Bertalanffy | 0.14 (0.016 - 0.67) [3] | -12.5 (-32 - 4.4) [5] | 1.19 [8] | 2096 [3] | 0.42 (0.16 - 1) [5] | 0.97 (0.81 - 1) | 100 | 3 |
| Exponential $V_0$ | 0.217 (0.0069 - 0.91) [6] | -10.7 (-34 - 5.1) [6] | -4.85 [3] | 2099 [4] | 0.49 (0.096 - 1.1) [6] | 0.93 (0.68 - 1) | 100 | 2 |
| Exponential-linear | 0.22 (0.048 - 0.76) [7] | -8.51 (-17 - 3.8) [7] | -2.7 [4] | 2174 [7] | 0.51 (0.27 - 1) [7] | 0.96 (0.91 - 0.99) | 100 | 2 |
| Logistic | 0.232 (0.05 - 0.73) [8] | -8.34 (-18 - 3.4) [8] | -2.52 [5] | 2214 [8] | 0.52 (0.27 - 0.98) [8] | 0.96 (0.92 - 0.99) | 100 | 2 |
| Exponential 1 | 1.36 (0.31 - 2.4) [9] | 6.01 (-5.4 - 13) [9] | 8.31 [9] | 2442 [9] | 1.2 (0.59 - 1.6) [9] | 0.64 (0.28 - 0.94) | 15 | 1 |

Models were ranked in ascending order of the $RMSE$, defined by expression (13). For each metric, indicated are the mean value (among all animals) and in parenthesis the minimal and maximal values (not reported for $AICc$ as they were redundant with the range of $AIC$). When reported, value inside brackets is the rank of the model for the underlying metric. The model ranking first is highlighted in bold. For animal $j$, $\frac{1}{I^j}\chi^2(\hat{\beta}^j)$ is the minimal value of the objective that was minimized in the individual fits approach (see (8)), divided by the number of time points $I^j$, and represents the variance of the weighted residuals. $AIC$ and $AICc$ are defined in (11) and (12), $AICc_{pop}$ is the $AICc$ resulting from the mixed-effect estimation (see Materials and Methods) and $R^2$ is defined in (14). Values reported in the $p$ column are percentages of animals were Kolmogorov-Smirnov test for normality of residuals was not rejected at the significance level of 0.05. # = number of parameters. $J = 20$ animals.

Table 2: Fit performances of the tumor growth models for the breast data ($J = 34$ animals)

| Model | $\frac{1}{I}\chi^2$ | AIC | AICc | AICc$_{pop}$ | RMSE | $R^2$ | p > 0.05 | # |
|---|---|---|---|---|---|---|---|---|
| Exponential-linear | 0.0919 (0.016 - 0.49) [2] | **-11.7 (-25 - 1) [1]** | **0.798 [1]** | **2832 [1]** | **0.34 (0.16 - 0.83) [1]** | 0.92 (0.66 - 0.99) | 100 | 2 |
| Gompertz | 0.0976 (0.015 - 0.33) [4] | -11.3 (-28 - -0.85) [2] | 1.21 [2] | 2866 [3] | 0.35 (0.14 - 0.68) [2] | 0.92 (0.67 - 0.99) | 100 | 2 |
| Generalized logistic | **0.0814 (0.0037 - 0.33) [1]** | -10.7 (-26 - 0.19) [4] | 11.9 [6] | 2870 [5] | 0.36 (0.096 - 0.76) [3] | **0.94 (0.8 - 0.99)** | 100 | 3 |
| Power law | 0.102 (0.016 - 0.32) [5] | -10.9 (-22 - -0.017) [3] | 1.53 [3] | 2913 [7] | 0.36 (0.16 - 0.71) [4] | 0.92 (0.61 - 0.99) | 100 | 2 |
| Exponential $V_0$ | 0.118 (0.011 - 0.37) [7] | -9.86 (-20 - 1) [5] | 2.61 [4] | 2870 [4] | 0.39 (0.13 - 0.78) [5] | 0.9 (0.56 - 0.99) | 100 | 2 |
| Von Bertalanffy | 0.0928 (0.015 - 0.32) [3] | -9.77 (-26 - 1.2) [6] | 11.9 [7] | 2876 [6] | 0.39 (0.15 - 0.8) [6] | 0.93 (0.67 - 0.99) | 100 | 3 |
| Dynamic CC | 0.11 (0.018 - 0.5) [6] | -8.58 (-20 - 3.2) [7] | 13 [9] | 2862 [2] | 0.42 (0.21 - 0.94) [7] | 0.91 (0.58 - 0.99) | 100 | 3 |
| Logistic | 0.145 (0.0037 - 0.42) [8] | -8.1 (-22 - -0.13) [8] | 4.37 [5] | 2921 [8] | 0.43 (0.078 - 0.76) [8] | 0.86 (0.65 - 0.99) | 100 | 2 |
| Exponential 1 | 2.19 (0.62 - 3.4) [9] | 8.77 (0.62 - 13) [9] | 12.6 [8] | 3518 [9] | 1.6 (0.85 - 2) [9] | -0.91 (-5.9 - 0.88) | 53 | 1 |

Table 3: Parameter values estimated from the fits. Lung data

| Model | Par. | Unit | Median value (CV) | Mean normalized std error (CV) |
|---|---|---|---|---|
| Power law | $a$ | $\left[mm^{3(1-\gamma)} \cdot day^{-1}\right]$ | 0.921 (38.9) | 11.9 (48.7) |
| | $\gamma$ | - | 0.788 (9.41) | 4 (53.4) |
| Gompertz | $a$ | $[day^{-1}]$ | 0.742 (25.3) | 6.02 (51.3) |
| | $\beta$ | $[day^{-1}]$ | 0.0792 (42.4) | 13.7 (65.4) |
| Exponential-linear | $a_0$ | $[day^{-1}]$ | 0.49 (19.3) | 3.08 (41.5) |
| | $a_1$ | $[mm^3 \cdot day^{-1}]$ | 115.6 (22.6) | 15.7 (40.7) |
| Dynamic CC | $a$ | $[day^{-1}]$ | 0.399 (106) | 447 (89.8) |
| | $b$ | $[mm^{-2} \cdot day^{-1}]$ | 2.66 (241) | 395 (176) |
| | $K_0$ | $[mm^3]$ | 2.6 (322) | 6.5e+04 (345) |
| Von Bertalanffy | $a$ | $\left[mm^{3(1-\gamma)} \cdot day^{-1}\right]$ | 7.72 (112) | 1.43e+04 (155) |
| | $\gamma$ | - | 0.947 (13.5) | 40.9 (73) |
| | $b$ | $[day^{-1}]$ | 6.75 (118) | 2.98e+07 (222) |
| Generalized logistic | $a$ | $[day^{-1}]$ | 2555 (148) | 2.36e+05 (137) |
| | $K$ | $[mm^3]$ | 4378 (307) | 165 (220) |
| | $\alpha$ | - | 0.0001413 (199) | 2.36e+05 (137) |
| Exponential $V_0$ | $V_0$ | $[mm^3]$ | 13.2 (47.9) | 28.9 (55) |
| | $a$ | $[day^{-1}]$ | 0.257 (15.4) | 7.49 (48.3) |
| Logistic | $a$ | $[day^{-1}]$ | 0.502 (17.5) | 3.03 (48.9) |
| | $K$ | $[mm^3]$ | 1297 (23.1) | 17.2 (43.8) |
| Exponential 1 | $a$ | $[day^{-1}]$ | 0.399 (13.8) | 2.87 (24.5) |

Shown are the median values within the population and in parenthesis the coefficient of variation (CV, expressed in percent and defined as the standard deviation within the population divided by mean and multiplied by 100) that quantifies inter-animal variability. Last column represents the normalized standard errors ($nse$) of the maximum likelihood estimator, defined in (11).

Table 4: **Parameter values estimated from the fits. Breast data**

| Model | Par. | Unit | Median value (CV) | Mean normalized std error (CV) |
|---|---|---|---|---|
| Power law | $a$ | $\left[mm^{3(1-\gamma)} \cdot day^{-1}\right]$ | 1.32 (74.1) | 31.2 (48.6) |
| | $\gamma$ | - | 0.58 (23) | 12.1 (62.2) |
| Gompertz | $a$ | $[day^{-1}]$ | 0.56 (18.4) | 7.52 (43) |
| | $\beta$ | $[day^{-1}]$ | 0.0719 (26.4) | 12.5 (65.5) |
| Exponential-linear | $a_0$ | $[day^{-1}]$ | 0.31 (16.8) | 6.22 (65.9) |
| | $a_1$ | $[mm^3 \cdot day^{-1}]$ | 67.8 (33.2) | 12.9 (45.4) |
| Dynamic CC | $a$ | $[day^{-1}]$ | 2.63 (81.3) | 597 (339) |
| | $b$ | $[mm^{-2} \cdot day^{-1}]$ | 0.829 (399) | 1.33e+03 (571) |
| | $K_0$ | $[mm^3]$ | 12.7 (525) | 6.48e+03 (361) |
| Von Bertalanffy | $a$ | $\left[mm^{3(1-\gamma)} \cdot day^{-1}\right]$ | 2.32 (113) | 1.17e+04 (181) |
| | $\gamma$ | - | 0.918 (22.5) | 128 (65.5) |
| | $b$ | $[day^{-1}]$ | 0.808 (132) | 1.48e+08 (300) |
| Generalized logistic | $a$ | $[day^{-1}]$ | 2753 (131) | 7.41e+05 (160) |
| | $K$ | $[mm^3]$ | 1964 (557) | 232 (433) |
| | $\alpha$ | - | 2.675e-05 (166) | 7.41e+05 (160) |
| Exponential $V_0$ | $V_0$ | $[mm^3]$ | 68.2 (57.2) | 34.5 (50.8) |
| | $a$ | $[day^{-1}]$ | 0.0846 (27.7) | 13.7 (44.1) |
| Logistic | $a$ | $[day^{-1}]$ | 0.305 (10.2) | 3.17 (34.9) |
| | $K$ | $[mm^3]$ | 1221 (31.4) | 11.8 (73.8) |
| Exponential 1 | $a$ | $[day^{-1}]$ | 0.223 (5.9) | 3.72 (21.3) |

Shown are the median values within the population and in parenthesis the coefficient of variation (CV, expressed in percent and defined as the standard deviation within the population divided by mean and multiplied by 100) that quantifies inter-animal variability. Last column represents the normalized standard errors ($nse$) of the maximum likelihood estimator, defined in (11).

**Table 5: Predictive power. Lung data**

| Model | Overall mean success | $S_{5,glob}$ | $S_{5,2}$ | $RE_{5,2}$ | $S_{3,1}$ | $RE_{3,1}$ | $S^f_{3,1}$ | $RE^f_{3,1}$ |
|---|---|---|---|---|---|---|---|---|
| Von Bertalanffy | 44.5 | 9/20 | 7/11 | 0.19 (0.016 - 0.51) | 7/14 | 0.29 (0.011 - 1.08) | 15/16 [87.5] | 0.10 (0.007 - 0.26) |
| Dynamic CC | 42.0 | 7/20 | 7/11 | 0.20 (0.010 - 0.48) | 7/14 | 0.27 (0.002 - 1.01) | 13/16 [62.5] | 0.14 (0.012 - 0.65) |
| Power law | 42.0 | 7/20 | 6/11 | 0.21 (0.019 - 0.52) | 8/14 | 0.29 (0.011 - 1.08) | 15/16 [64.1] | 0.08 (0.007 - 0.29) |
| Gompertz | 41.5 | 9/20 | 7/11 | 0.25 (0.069 - 0.57) | 6/14 | 0.30 (0.030 - 1.08) | 15/16 [119] | 0.10 (0.001 - 0.30) |
| Exponential $V_0$ | 39.3 | 6/20 | 6/11 | 0.33 (0.039 - 1.48) | 9/14 | 0.31 (0.035 - 1.20) | 15/16 [45.8] | 0.09 (0.000 - 0.27) |
| Exponential linear | 36.0 | 8/20 | 5/11 | 0.22 (0.037 - 0.43) | 10/14 | 0.21 (0.029 - 0.99) | 15/16 [31.2] | 0.10 (0.010 - 0.33) |
| Generalized logistic | 33.9 | 5/20 | 5/11 | 0.28 (0.069 - 0.57) | 5/14 | 0.31 (0.030 - 1.08) | 12/16 [110] | 0.14 (0.006 - 0.30) |

Models are presented in descending order of overall mean success (defined in (18)). $S_{n,d}$, defined in (17), is the success score for prediction when using n data points and predicting at future depth $d$, i.e. time $t_{n+d}$ (see Materials and Methods). For relative errors (20), mean value among animals is reported with ranges in parenthesis. $S^f_{3,1}$ and $RE^f_{3,1}$ stand for the success rates and relative errors for predictions of the late phase (see text for details). Reported in brackets in the $S^f_{3,1}$ column are the percent increase between $S_{3,1}$ and $S^f_{3,1}$.

**Table 6: Predictive power. Breast data**

| Model | Overall mean success | $S_{5,glob}$ | $S_{5,2}$ | $RE_{5,2}$ | $S_{3,2}$ | $RE_{3,2}$ | $S_{3,2}^f$ | $RE_{3,2}^f$ |
|---|---|---|---|---|---|---|---|---|
| Exponential linear | 83.8 | 20/25 | 17/23 | 0.13 (0.014 - 0.36) | 4/7 | 0.27 (0.079 - 0.84) | 17/20 [49] | 0.10 (0.001 - 0.32) |
| Dynamic CC | 63.3 | 20/25 | 19/23 | 0.14 (0.013 - 0.53) | 4/7 | 0.28 (0.071 - 0.87) | 13/20 [14] | 0.18 (0.001 - 0.43) |
| Power law | 62.3 | 18/25 | 17/23 | 0.15 (0.028 - 0.57) | 3/7 | 0.33 (0.092 - 0.97) | 14/20 [63] | 0.14 (0.001 - 0.41) |
| Gompertz | 59.0 | 18/25 | 17/23 | 0.15 (0.001 - 0.54) | 3/7 | 0.33 (0.076 - 0.97) | 13/20 [52] | 0.17 (0.001 - 0.61) |
| Von Bertalanffy | 58.8 | 18/25 | 17/23 | 0.15 (0.008 - 0.54) | 3/7 | 0.31 (0.077 - 0.87) | 14/20 [63] | 0.14 (0.007 - 0.43) |
| Exponential V0 | 47.7 | 14/25 | 17/23 | 0.16 (0.003 - 0.66) | 3/7 | 0.37 (0.065 - 1.24) | 16/20 [87] | 0.13 (0.024 - 0.37) |
| Generalized logistic | 34.2 | 13/25 | 15/23 | 0.18 (0.001 - 0.54) | 3/7 | 0.27 (0.100 - 0.73) | 14/20 [63] | 0.15 (0.001 - 0.41) |

Models are presented in descending order of overall mean success (defined in (18)). $S_{n,d}$, defined in (17), is the success score for prediction when using n data points and predicting at future depth $d$, i.e. time $t_{n+d}$ (see Materials and Methods). For relative errors (20), mean value among animals is reported with ranges in parenthesis. $S_{3,2}^f$ and $RE_{3,2}^f$ stand for the success rates and relative errors for predictions of the late phase (see text for details). Reported in brackets in the $S_{3,2}^f$ column are the percent increase between $S_{3,2}$ and $S_{3,2}^f$.

**Figure 1**: **Volume measurement error**. A. First measured volume $y_1$ against second one $y_2$. Also plotted is the regression line (correlation coefficient r = 0.98, slope of the regression = 0.96). B. Error $y_1 - y_m$ against approximation of the volume given by the average of the two measurement $y_m = (y_1+y_2)/2$. The $\chi^2$ test rejected Gaussian distribution of constant variance (p = 0.004) C. Histogram of the normalized error $(y_1-y_m)/E_m$ applying the error model given by (22) with α = 0.84 and $V_m$ = 83 mm$^3$. It shows Gaussian distribution (p = 0.196, $\chi^2$ test).

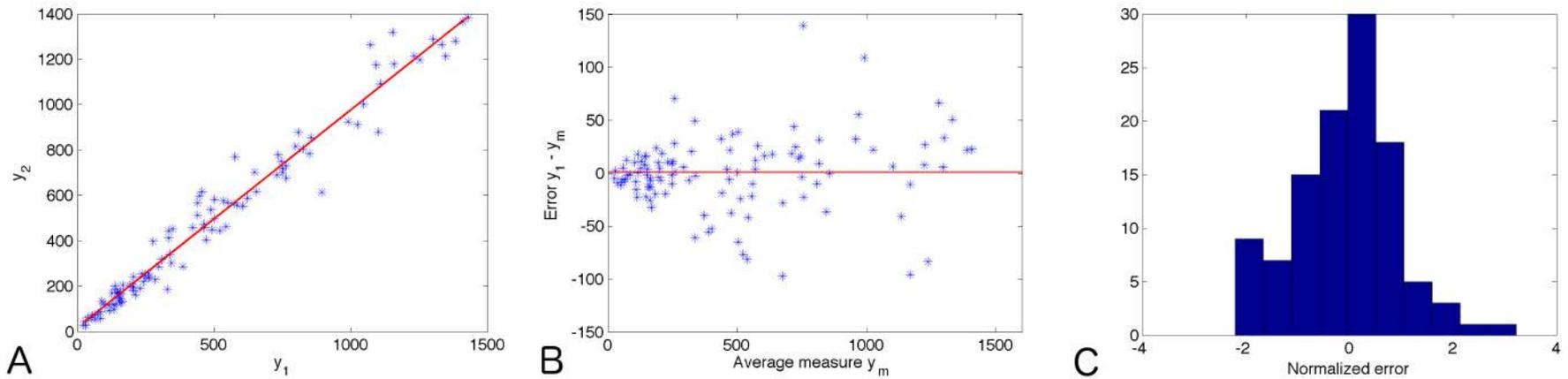

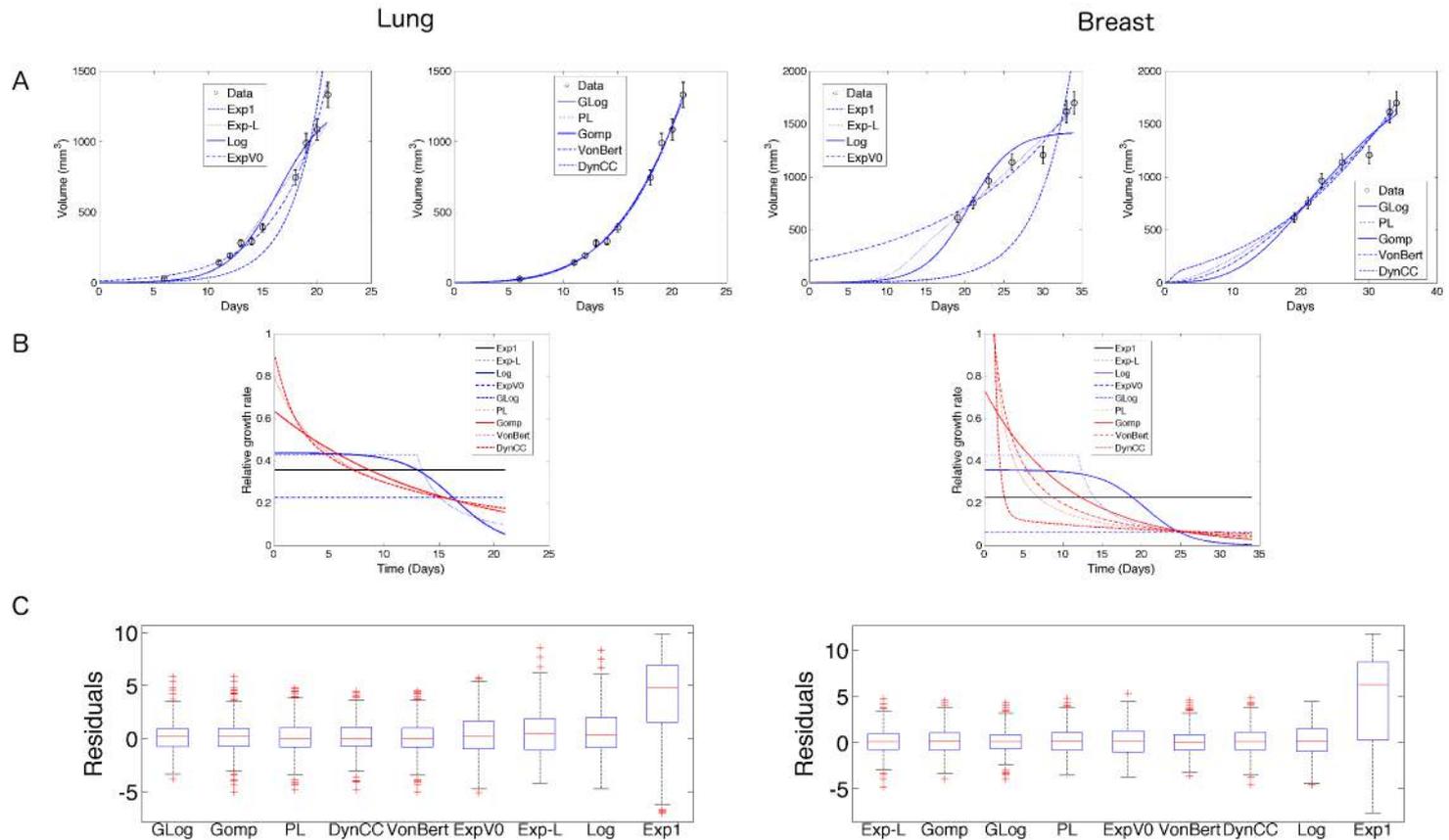

**Figure 2: Descriptive power of the models.** A. Representative examples of all growth models fitting the same growth curve (animal 10 for lung, animal 14 for breast). Errorbars correspond to the standard deviation of the *a priori* estimate of measurement error. In the lung setting, curves of the Gompertz, power law, dynamic CC and von Bertalanffy models are visually undistiguinshable. B. Corresponding relative growth rate curves. Curves for von Bertalanffy and power law are identical in the lung setting C. Residuals distributions, in ascending order of mean *RMSE* (13) over all animals. Residuals (see formula (15) for their definition) include fits over all the animals and all the time points. Exp1 = exponential 1, Exp-L = exponential-linear, Exp = exponential , Log = logistic, GLog = generalized logistic, PL = power law, Gomp= Gompertz, VonBert = von Bertalanffy, DynCC = dynamic CC.

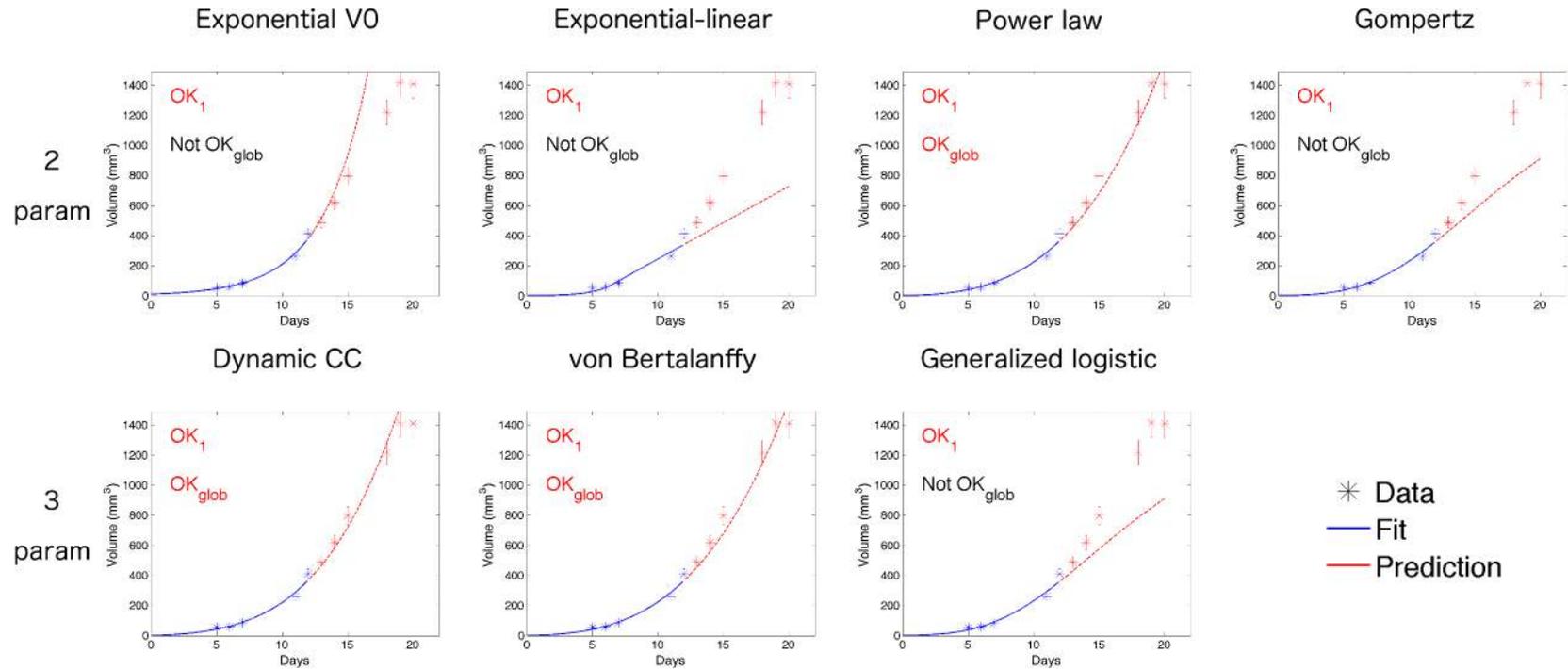

**Figure 3: Example of predictive power**. Representative example of the forecast performances of the models for the lung data set (mouse number 2). Five data points were used to estimate the animal parameters and predict future growth. Prediction success of the models are reported for the next day data point ($OK_1$) or global future curve ($OK_{glob}$), based on the criterion of a normalized error smaller than 3 (meaning that the median model prediction is within 3 standard deviations of the measurement error) for $OK_1$ and the median of this metric over the future curve for $OK_{glob}$ (see Materials and Methods for details).

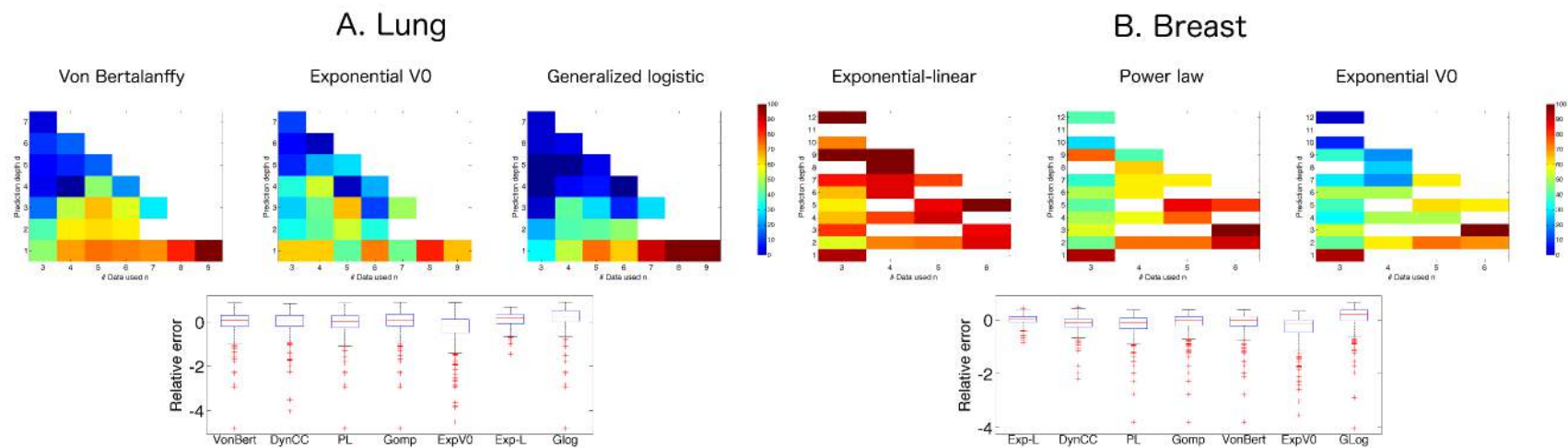

**Figure 4: Prediction depth and number of data points**. Predictive power of some representative models depending on the number of data points used for estimation of the parameters (*n*) and the prediction depth in the future (*d*). Top: at position the color represents percentage of successfully predicted animals when using data points and forecasting the time point $t_n+d$, i.e. the score $S_{n,d}$ (multiplied by 100), defined in (17). This proportion only includes animals having measurements at these two time points, thus values at different rows on the same column or reverse might represent predictions in different animals. White squares correspond to situations where this number was too low (<5) and thus success score, considered not significative, was not reported. Bottom: distribution of the relative error of prediction, all animals and *(n,d)* settings pooled together. Models were ranked in ascending order of overall mean success score reported in Tables 5 and 6. A. Lung tumor data. B. Breast tumor data.

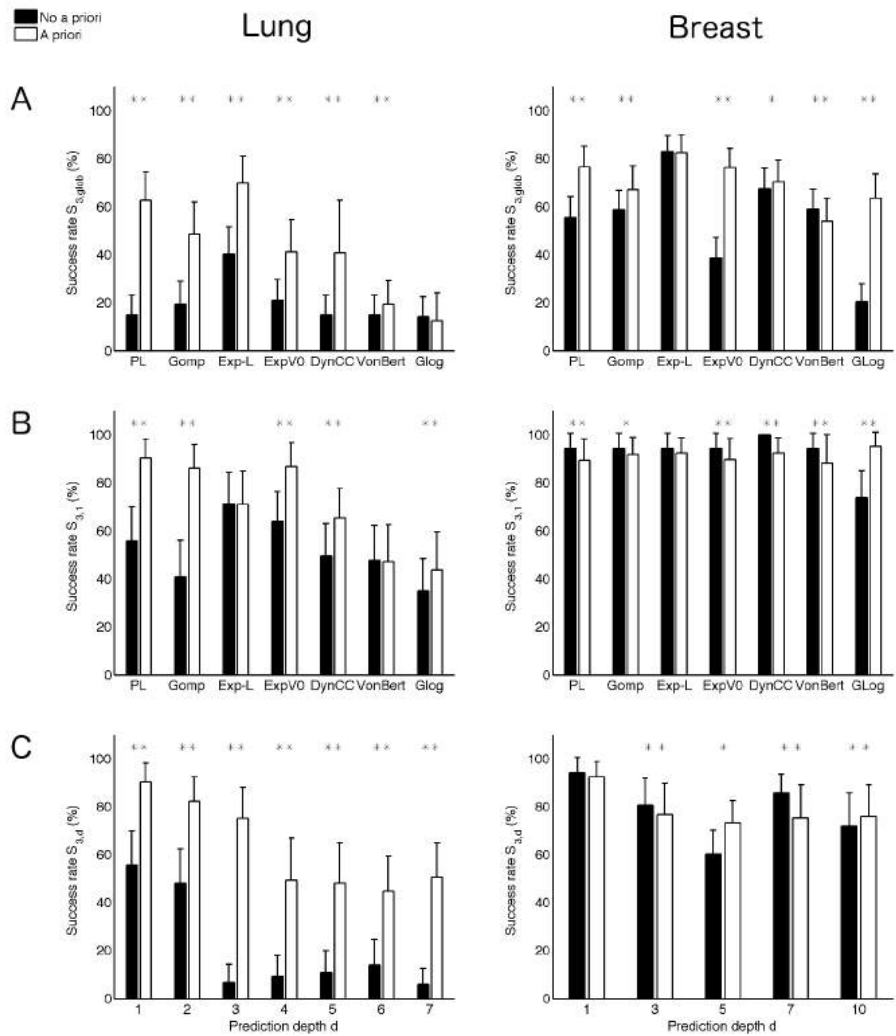

**Figure 5: *A priori* information and improvement of prediction success rates**. Predictions were considered when randomly dividing the animals between two equal groups, one used for learning the parameters distribution and the other for prediction, using *n=3* data points. Success rates are reported as mean ± standard deviation over 100 random partitions into two groups. Prediction of global future curve, quantified by the score $S_{3,glob}$ (see Materials and Methods). B. Benefit of the method for prediction of the next day, using three data points (score $S_{3,1}$). C. Prediction improvement at various prediction depths, using the power law model (lung data) or the exponential-linear model (breast data). Due to lack of animals to be predicted for some of the random assignments, results of depths 2, 4, 6 and 9 for the breast data were not considered significant (see Materials and Methods).
*= *p < 0.05, \*\* = p < 0.001,* Student's t-test

# Text S1: Numerical procedures for parameters estimation

When available, we used the analytical formula of the model for numerical computations. This was the case for all the models except the dynamic CC model (5), for which we used the ode solver *ode45* of Matlab for computation of the solution.

Several optimization algorithms implemented in Matlab [49] were preliminary tested for minimization of $\chi^2$ and the function *lsqcurvefit* (with trust-region algorithm and maximum number of iterations and function evaluations allowed both set to $10^5$) was decided to be well-suited for most of the settings based on its rapidity of convergence and ability to manage bounds constraints. Convergence of the algorithm was systematically checked and was effective for all the parameter estimations performed (for both the descriptive and predictive analysis). Values of the parameters used for initialization of the minimization algorithm were fixed from preliminary explorative analysis, sometimes with the help of the more global optimizer *fminsearch*. They are reported in the Table S1 and a short study of the impact of the initial guess on the values of the parameters resulting from the optimization is provided in the supplementary text S2 and supplementary Table S3.

We compared the fits obtained by *lsqcurvefit* to other regression functions implemented in Matlab, namely *nlinfit* (Levenberg-Marquardt algorithm) or *fminsearch* (Nelder-Mead algorithm). For the identifiable models (in the sense of the practical identifiability reported in text S2), as well as for the dynamic CC model, no significant differences in the estimated parameter sets were observed (Student's t-test) when comparing fits obtained with either *nlinfit* or *fminsearch* to the fits of *lsqcurvefit*. For instance, the relative difference for parameters estimates of the full lung data set with the power law model was lower than $1.5 \times 10^{-4}$ and $p$ values of Student's t-test for significant differences between the population distributions of parameter sets were all larger than 0.99. On the other hand, consistently with their dependence on the initialization of the optimization algorithm (Table S2), parameters estimations of the two non-identifiable models generalized logistic (3) and von Bertalanffy (6), were significantly sensitive to the algorithm used. However, for the von Bertalanffy model, this difference in the parameters did not impact on the resulting growth curve and generated the same fits (identical value of the minimal sum of squared errors). This observation suggests a structural non-identifiability of this model for the range of the observed dynamics, that is, that different parameter values can generate the same curve.

Two models required specific attention, namely the generalized logistic model (3) and the exponential-linear model (1). In the first setting, due to low identifiability, *lsqcurvefit* was found to converge to irrealistic local minima and we rather used the more global optimizer *fminsearch* with an increased maximum number of iterations and function evaluations allowed (from 600 to $10^5$), in order to ensure convergence of the algorithm. Convergence was checked *a posteriori* to be effective. For the exponential-linear model, local minima issues were also likely to happen because, on a finite time-course and within some parameter ranges, the model could be insensitive to variation of one of the parameter and remain in a fully-linear or fully-exponential phase (if the resulting $\tau$ was larger than the maximum time or lower than the minimal one), which could be sub-optimal. A useful workaround was to impose upper bounds on the parameters (in particular $a_1$). To avoid numerical issues when solving the von Bertalanffy and power law models, we also imposed an upper bound on parameter $\gamma$. In the setting of *lsqcurvefit*, this required imposition of a bound also on parameters $a$ and $b$, which were taken large enough not to be active.

# Text S2: Practical identifiability of the models

Sensitivity of the fits obtained from our estimation procedure was assessed for each model by systematically varying initialization of the algorithm. Only the LLC tumor data set was used for this study, in order to limit the computational cost and because the results should be similar with the LM2-4$^{LUC+}$ data. A compact subset of the parameter space of length 4 standard deviations was meshed and explored in each parameter direction above and below the mean population value from the fits reported in the text (Table 3). For the generalized logistic and von Bertalanffy models, due to high inter-animal variability of the parameters estimates (Table 3), this method led to extreme values of the parameters. These were not relevant initializations because they resulted in extreme, biologically unrealistic, behaviors of the models. Starting from these initializations, the minimization procedure was unable to generate good fits and consequently the identifiability scores were very low, but for a methodological reason and not due to structural properties of the models. Consequently, a different procedure was employed. The baseline mean value was taken to be the one of Table S1 and standard deviation was replaced by an arbitrary standard deviation of 50% of the baseline value. For each model, $20 \times 11^P$ individual fits were performed (for the 20 animals growth curves) with $P$ the number of parameters in the model, giving a total of 92180 individual tumor growth performed for the analysis.

We reported in Table S3 results of sensitivity scores. The first score was defined as the fraction of minimization runs (over all the animals and initializations) for which the minimized objective (defined in (8)) converged to the same value as when starting from the baseline initialization (within a 10% relative error). This score essentially verifies that all the resulting fits, even if converging to a different parameter vector than with the baseline initialization, were equally good. A low fit score would mean that the optimization algorithm has often converged to sub-optimal minima. All of the models had very good fit score (Table S3).

The second score concerned parametric identifiability and was defined as the fraction of minimization runs that converged to the same parameter vector as when performing minimization starting from the baseline value (within a 10% relative error). When global numerical identifiability was not observed, further investigation on each parameter was performed and the resulting variation of the best-fit parameter computed (Table S3). Low parametric identifiability scores means that the model was able to generate almost identical growth curves with different parameter sets.

As could be expected from their large standard errors on the parameters estimates (Table 3), the models with three parameters (von Bertalanffy (6), dynamic CC (5) and generalized logistic (3)) exhibited low parametric identifiability scores (of respectively 34.4%, 45.1% and 56.5%, out of 26620 fits for each). For the von Bertalanffy model, the non-identifiability was probably due to the absence of identifiable range of convergence to an asymptotic volume in our data. The low identifiability of the dynamic CC model can be explained by its two-dimensional nature (volume and carrying capacity are variables) with only one observable used for the fits. This fact resulted in variability mostly in estimation of $K_0$. For the generalized logistic model, non-identifiability mostly came from parameters $a$ and $\alpha$. This could be explained by the high flexibility of the model that generated several close-to-optimal fits, although with different parameter values. Most of the fits (93.6%) were in the range of 10% from the baseline fit (i.e. the fits used in Table 1). Considering a different parameterization of the model (changing $a$ into $\tilde{a} = \frac{a}{\alpha}$) might improve the identifiability of the generalized logistic model.

All the other models exhibited very good robustness in the parameter estimation.

**Table S1: Initializations of the least squares minimization algorithm.**

| Model | Parameter | Initialization | Upper bound |
|---|---|---|---|
| Power law | $a$ | 1 | 100 |
| | $\gamma$ | 2/3 | 1 |
| Gompertz | $a$ | 1 | - |
| | $\beta$ | 0.1 | - |
| Dynamic CC | $a$ | 3 | - |
| | $b$ | 0.5 | - |
| | $K_0$ | 10 | - |
| Generalized logistic | $a$ | 10 | - |
| | $K$ | 10000 | - |
| | $\alpha$ | 0.01 | - |
| Von Bertalanffy | $a$ | 1 | 100 |
| | $\gamma$ | 0.75 | 1 |
| | $b$ | 0.1 | 100 |
| Exponential $V_0$ | $a$ | 0.1 | - |
| | $V_0$ | 20 | - |
| Exponential-linear | $a_0$ | 0.2 | 10 |
| | $a_1$ | 100 | 200 |
| Logistic | $a$ | 1 | - |
| | $K$ | 10000 | - |
| Exponential 1 | $a$ | 0.1 | - |

Also reported are bounds used in *lsqcurvefit* for estimation of parameters of the power law, von Bertalanffy and exponential-linear models.

**Tables S2: Comparison of individual fits between the individual and population approaches.** Fits were performed using either an individual estimation of the growth curves based on weighted least-squares estimation or a population approach. In both settings, the error model was proportional to the volume to the power $\alpha = 0.84$. The only difference was that Monolix estimation did not allow for a setting with a threshold volume $V_m$, which was thus taken to be 0. However, due to its low value ($V_m = 83$ mm$^3$), it was not very active in the individual approach. Reported are the mean (over the time points) weighted least squares (i.e. the ones of the first column of Tables 1, 2 except with $V_m = 0$, i.e. the Monolix setting), for both approaches. More precisely, if

$$\frac{1}{I^j}\chi^2(\hat{\beta}^j_\phi) = \frac{1}{I^j}\sum_{i=1}^{I^j}\left(\frac{y_i^j - M(t_i^j, \hat{\beta}^j_\phi)}{\left(y_i^j\right)^\alpha}\right)^2, \quad \phi = \mathcal{I}, \mathcal{P}$$

with $\hat{\beta}^j_\phi$ being the individual estimate of parameter set $\beta$ in animal $j$, using either the individual approach ($\phi = \mathcal{I}$) or the population approach ($\phi = \mathcal{P}$). The numbers reported are the mean value of $\frac{1}{I^j}\chi^2(\hat{\beta}^j_\phi)$ (over the population, i.e. index $j$) as well as minimal and maximal values. Note that due to the relatively large volumes of the breast data, $V_m$ was not active and the values of the individual approach are exactly the ones of Table 2 in this case (S2.B). Last column is the $p$-value of Student's t-test for significant differences between the individual and population approaches.

### S2.A Lung data

| Model | Indiv. (least squares) | Pop. (Monolix) | $p$ |
|---|---|---|---|
| DynCC | 0.236(0.0143-2.33) | 0.515(0.0228-6.66) | 0.421 |
| Gomp | 0.274(0.0213-2.74) | 0.768(0.0422-10) | 0.335 |
| PL | 0.27(0.0176-2.68) | 0.988(0.0362-13.9) | 0.308 |
| VonBert | 0.256(0.0176-2.74) | 0.735(0.032-10.5) | 0.374 |
| ExpV0 | 0.373(0.00718-2.9) | 0.947(0.0638-12.4) | 0.36 |
| GLog | 0.186(0.0213-1.25) | 0.799(0.0478-10.3) | 0.234 |
| Log | 0.349(0.0547-1.81) | 0.977(0.12-4.38) | 0.0471 |
| Exp-L | 0.322(0.0529-1.32) | 0.614(0.0621-3) | 0.12 |
| Exp1 | 1.46(0.335-2.58) | 2.53(0.924-10.7) | 0.0686 |

**S2.B Breast data**

| Model | Indiv. (least squares) | Pop. (Monolix) | $p$ |
| --- | --- | --- | --- |
| DynCC | 0.11(0.018-0.503) | 0.25(0.0352-1.75) | 0.0448 |
| Gomp | 0.0976(0.0147-0.328) | 0.237(0.039-1.87) | 0.0306 |
| PL | 0.102(0.0159-0.323) | 0.216(0.0311-1.49) | 0.0367 |
| VonBert | 0.0928(0.0148-0.323) | 0.263(0.0337-2.11) | 0.0231 |
| ExpV0 | 0.118(0.0106-0.37) | 0.278(0.0121-2.17) | 0.0484 |
| GLog | 0.0814(0.00366-0.328) | 0.226(0.0361-1.8) | 0.0181 |
| Log | 0.145(0.00367-0.417) | 0.178(0.0234-0.66) | 0.232 |
| Exp-L | 0.0919(0.0159-0.49) | 0.113(0.0273-0.615) | 0.394 |
| Exp1 | 2.19(0.617-3.44) | 5.88(0.629-27.4) | 0.00157 |

**Table S3: Practical identifiability.** Two identifiability scores were reported. The fit score is the proportion of minimization runs, among the $20 \times N^P$ performed, for which the resulting minimized objective converged to the same value as when starting from the baseline value. The global parametric score is the proportion of minimization runs that converged to the same parameter vector, within a 10% relative error. When this last score was lower than 100%, further analysis was performed and the same score was computed for each parameter of the model. We also report their median relative deviation to the base value, in percent. Param. = Parameter. Dev. = Deviation.

| Model | Fit score (%) | Global param. score (%) | Param. score (%) (Median Dev.) | | |
|---|---|---|---|---|---|
| Power law | 100.0 | 100.0 | $a$: 100.0 (0.000811) | $\gamma$: 100.0 (0.000272) | |
| Gompertz | 100.0 | 100.0 | $a$: 100.0 (0.000273) | $\beta$: 100.0 (0.000662) | |
| Dynamic CC | 85.9 | 45.1 | $a$: 67.3 (0.948) | $b$: 70.3 (0.491) | $K_0$: 45.2 (37.5) |
| Von Bertalanffy | 99.4 | 34.4 | $a$: 77.9 (1.53) | $\gamma$: 99.3 (0.0181) | $b$: 34.4 (21.3) |
| Generalized logistic | 93.6 | 56.5 | $a$: 56.9 (4.92) | $K$: 93.4 (0.00149) | $\alpha$: 57.0 (5.01) |
| Exponential $V_0$ | 100.0 | 100.0 | $V_0$: 100.0 (0.00134) | $a$: 100.0 (0.000366) | |
| Logistic | 97.7 | 97.7 | $a$: 97.7 (0.001) | $K$: 97.7 (0.00557) | |
| Exponential-linear | 95.0 | 97.3 | $a_0$: 97.3 (0.000517) | $a_1$: 97.3 (0.00262) | |
| Exponential 1 | 100.0 | 100.0 | $a$: 100.0 (0.00013) | | |

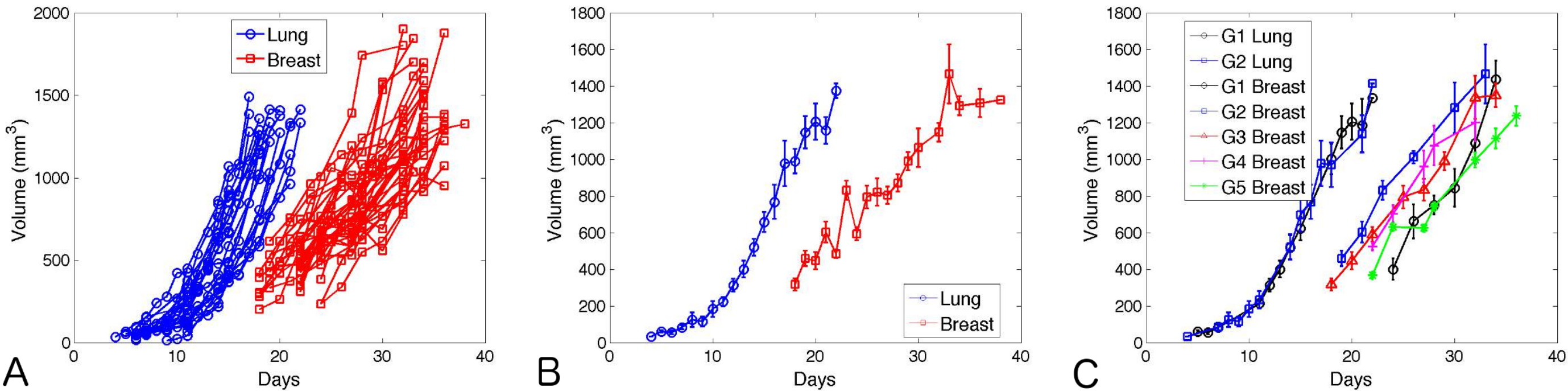

**Figure S1: Data.** Plots of the brute data sets from the lung and breast experiments. A. All animals growth curves. B. Average curves. C. Per group averages (lung and breast data resulted from combinations of respectively two and three separate experiments). G = group.

**Figure S2: Examples of individual predictions.** Lung data. Prediction success of the model are reported for the next day ($OK_1$) or global future curve ($OK_{glob}$), based on the criterion of a normalized error smaller than 3 (meaning that the median model prediction is within 3 standard deviations of the measurement error) for $OK_1$ and the median of this metric over the future curve for $OK_{glob}$. Future growth was predicted using 5 data points. Plotted here are results from the Gompertz model.

**A. Gompertz model (similar to von Bertalanffy, dynamic CC and power law models)**

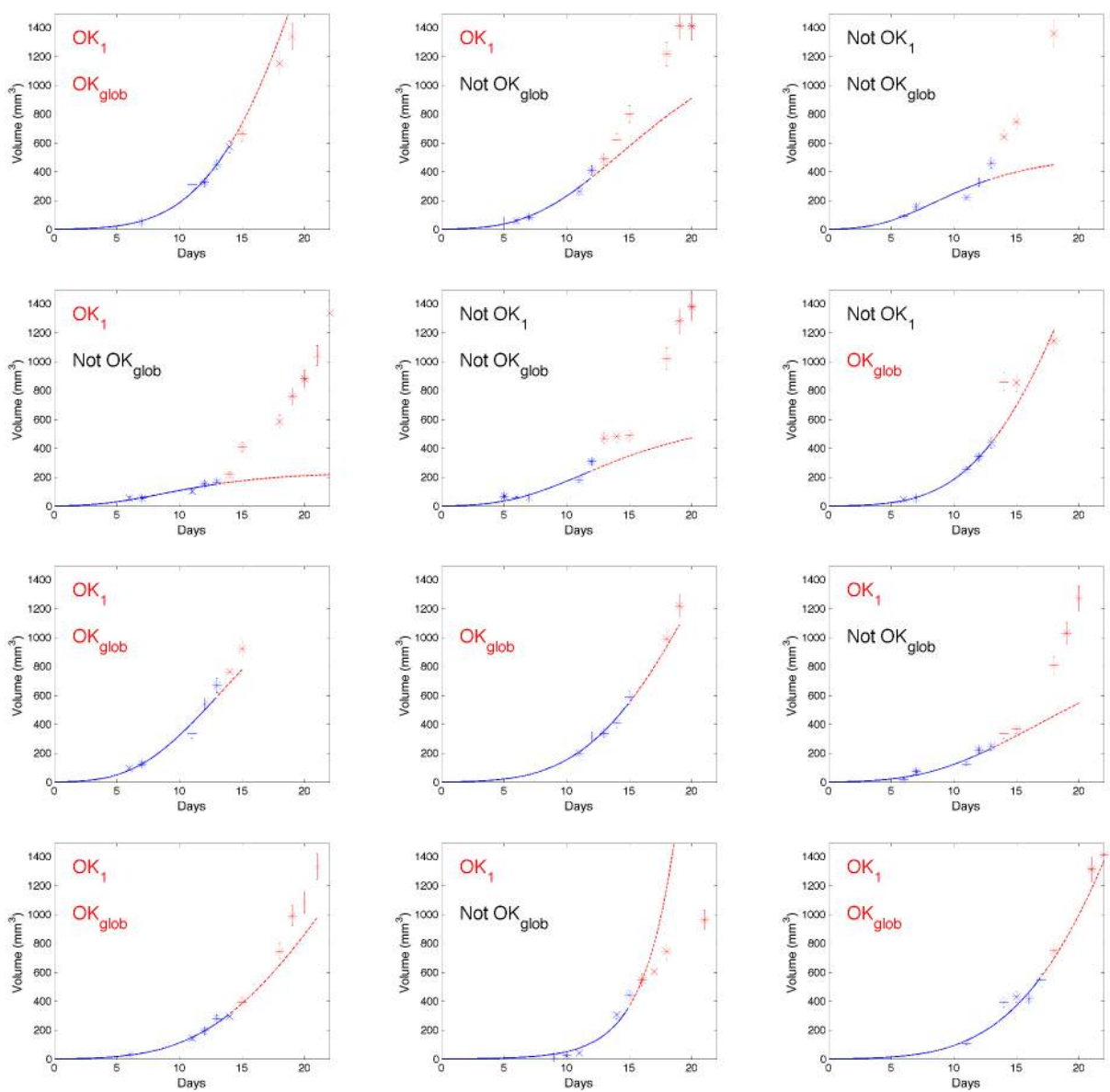

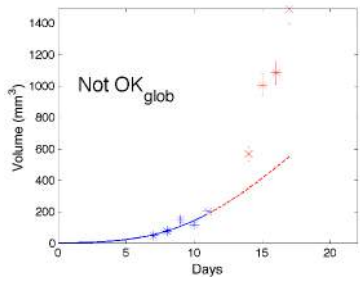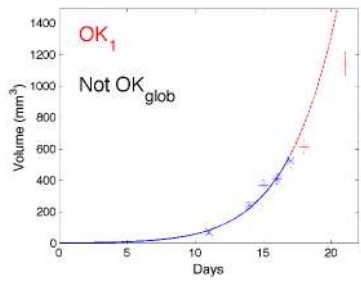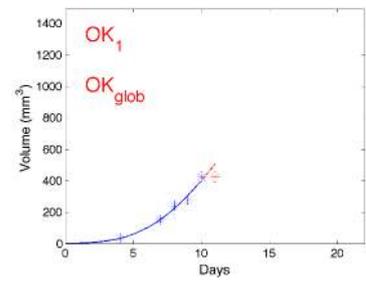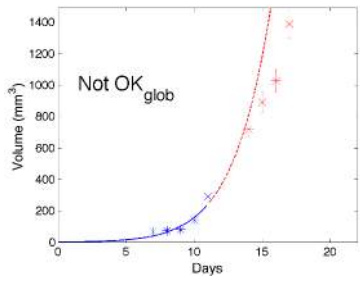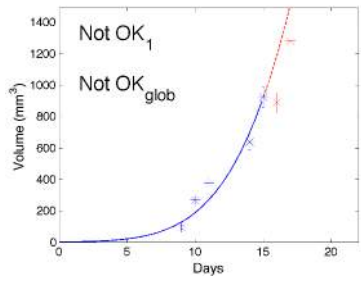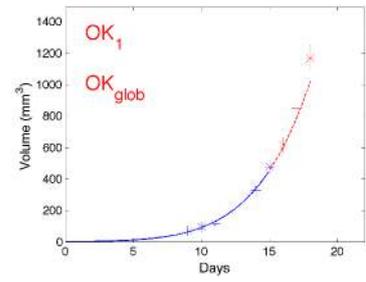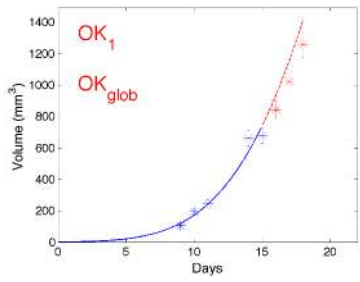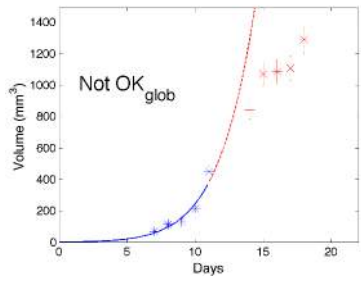

**B. Examples of sharp saturation of the generalized logistic model**

Mouse 12

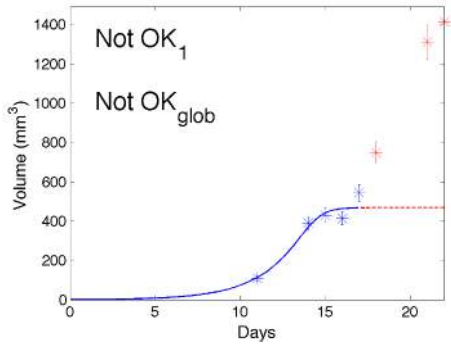

Mouse 14

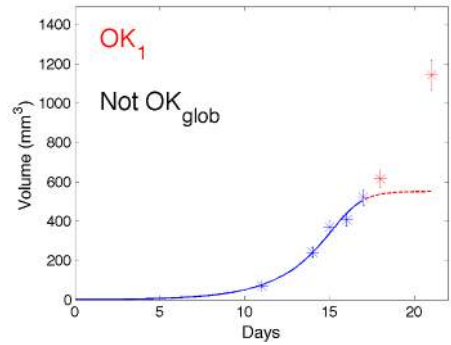

Mouse 17

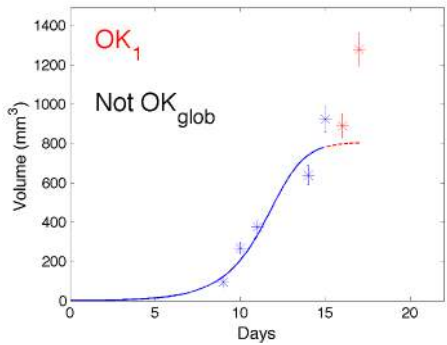

Mouse 19

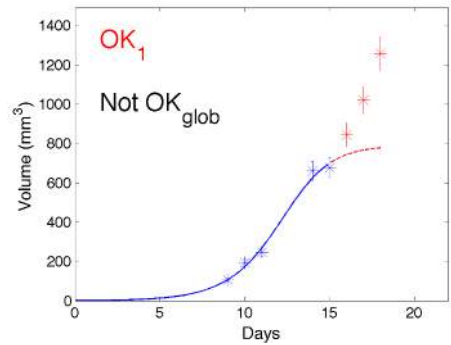

**Figure S3: Examples of individual predictions. Breast data.** Prediction success of the model are reported for the second next day data point ($OK_2$) or global future curve ($OK_{glob}$), based on the criterion of a normalized error smaller than 3 (meaning that the median model prediction is within 3 standard deviations of the measurement error) for $OK_2$ and the median of this metric over the future curve for $OK_{glob}$. Future growth was predicted using 5 data points and the exponential-linear model.

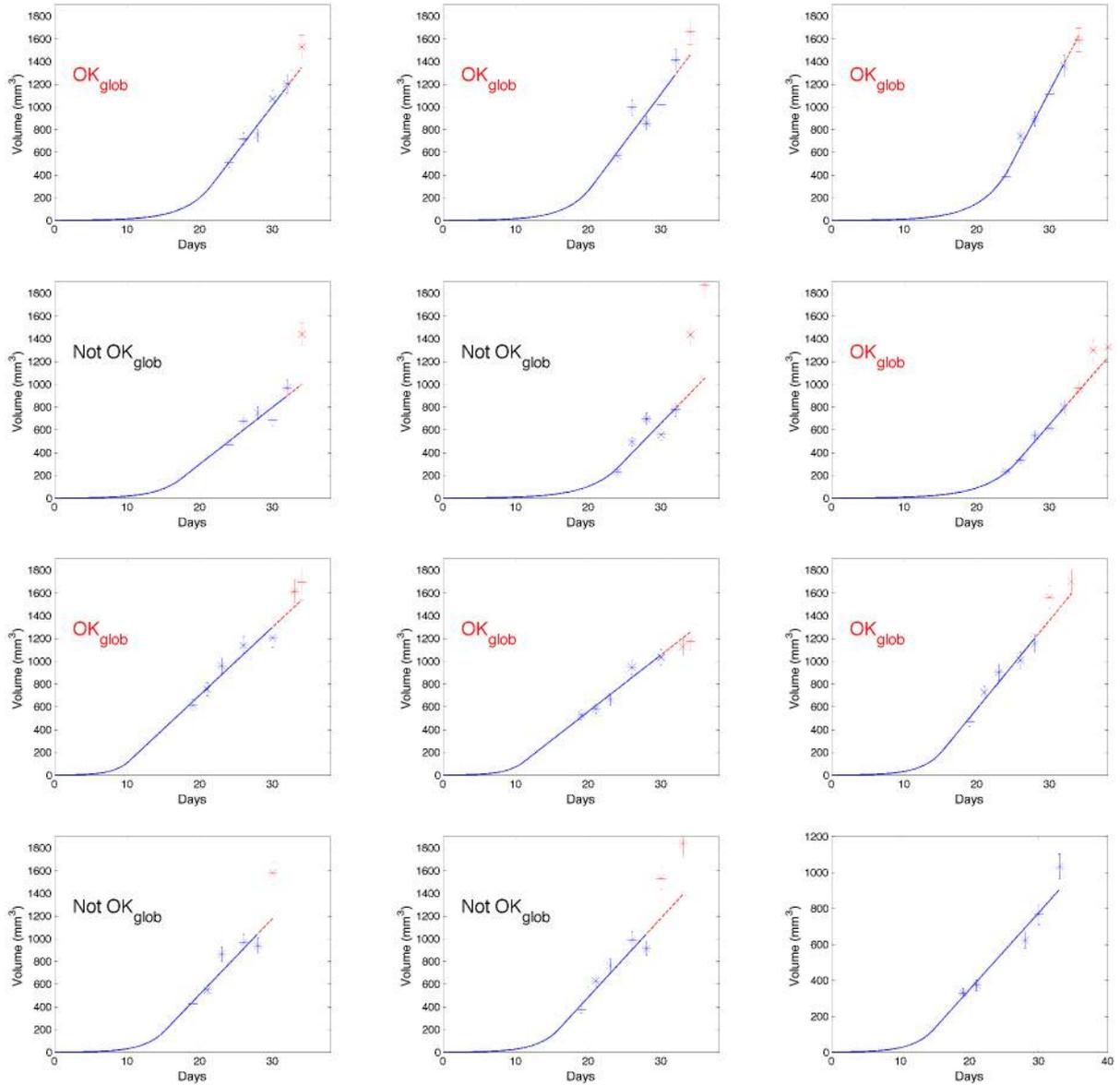

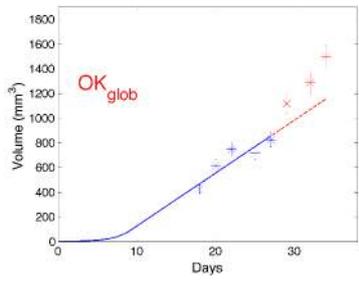
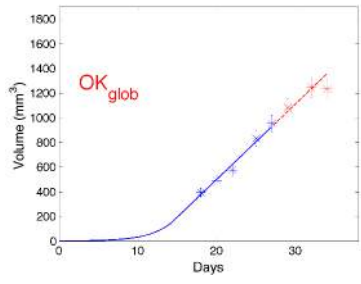
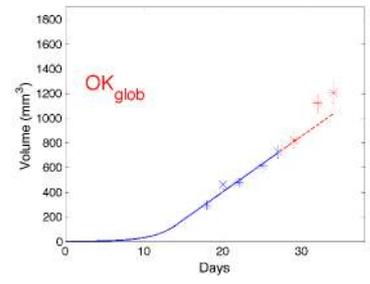
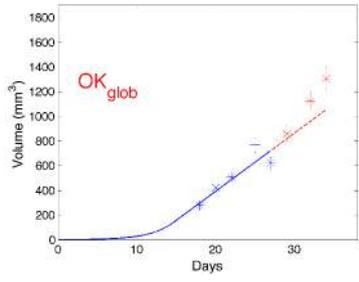
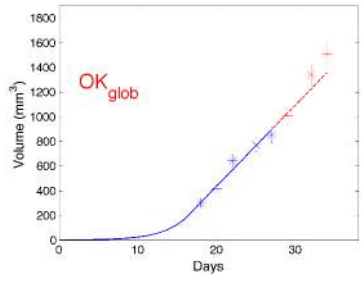
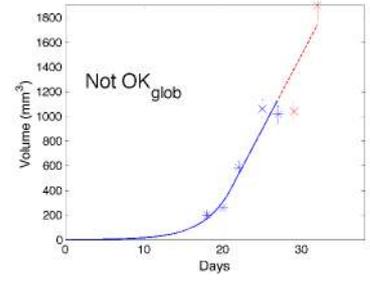
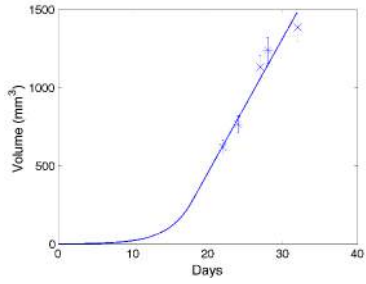
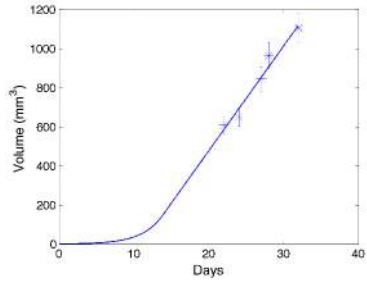
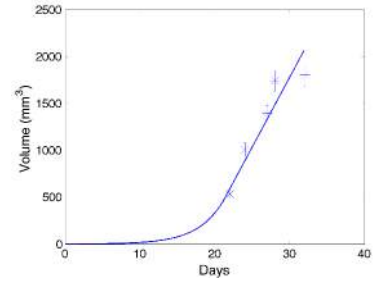
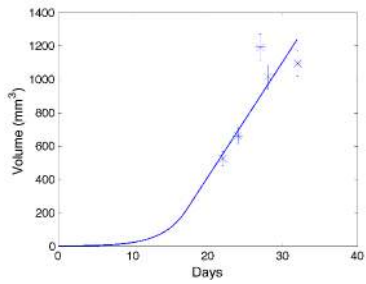
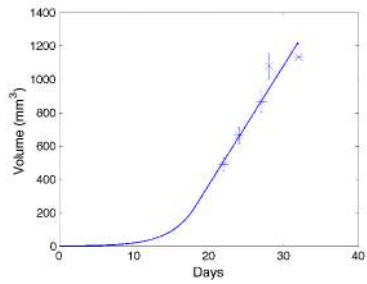
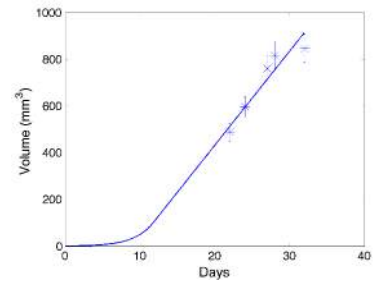

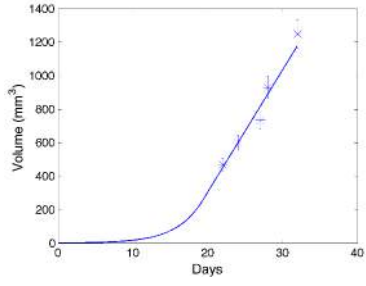
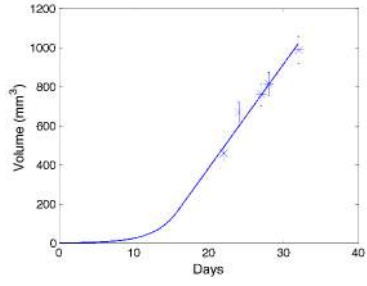
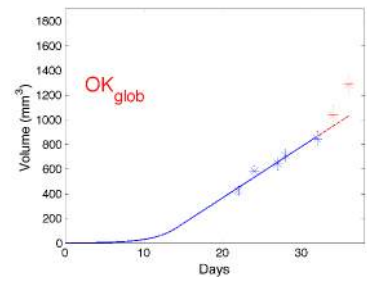
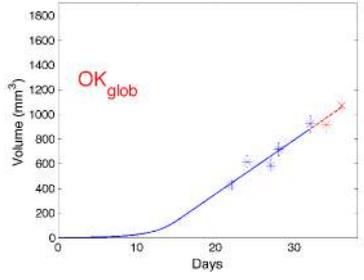
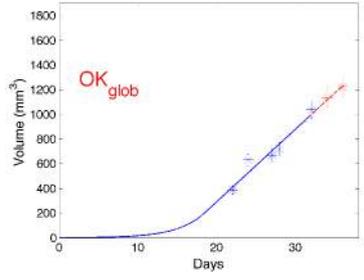
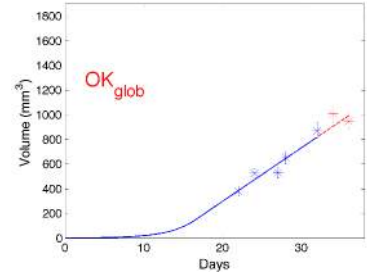
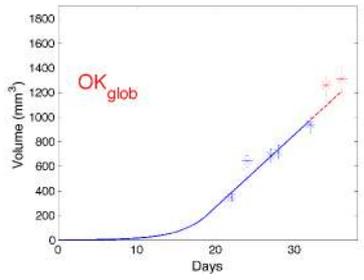
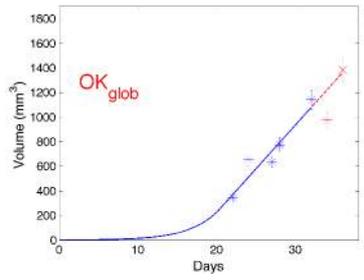
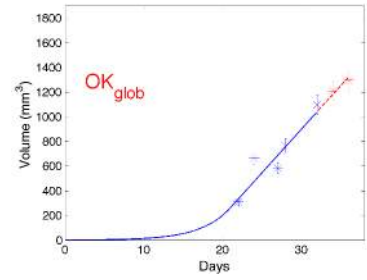
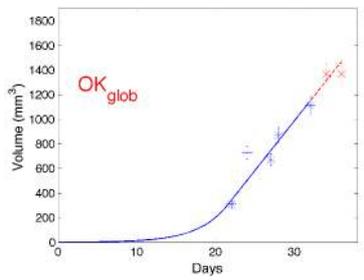

# A. Lung

### Dynamic CC
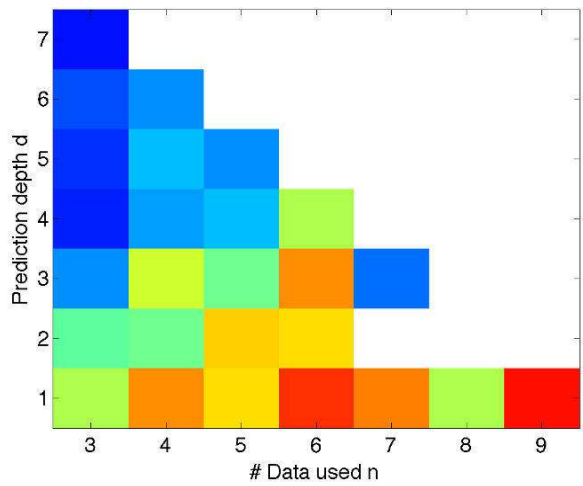

### Gompertz
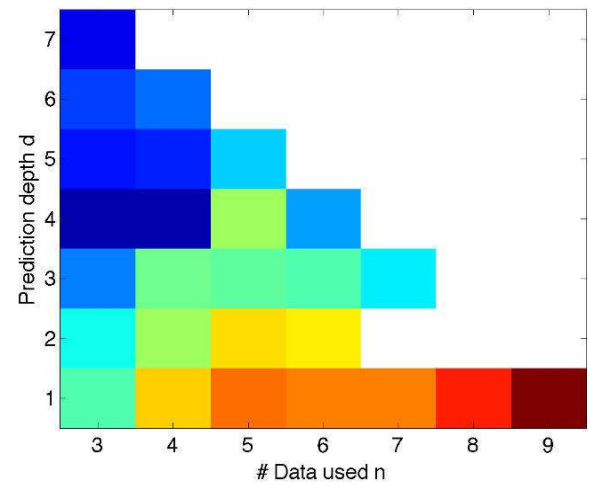

# B. Breast

### Generalized logistic
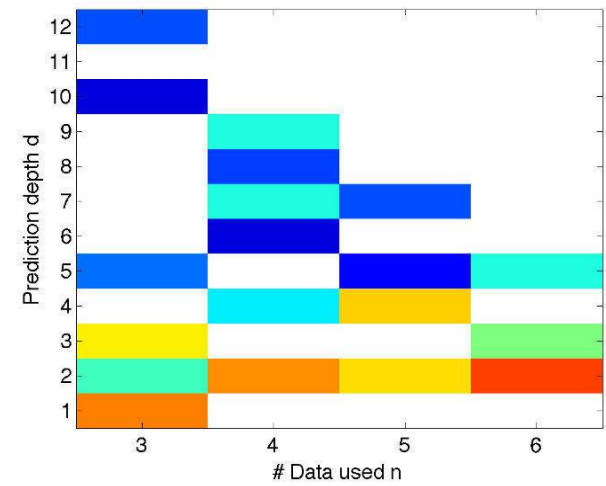

### Dynamic CC
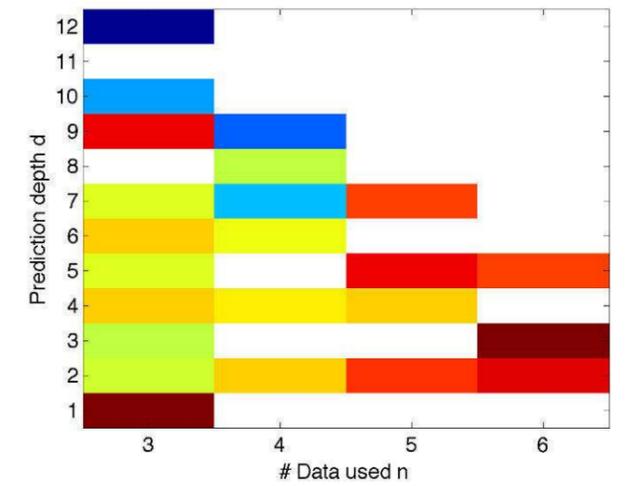

### Power law
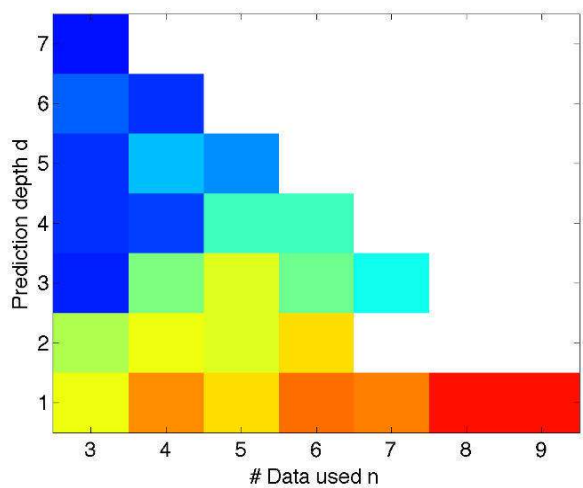

### Exponential linear
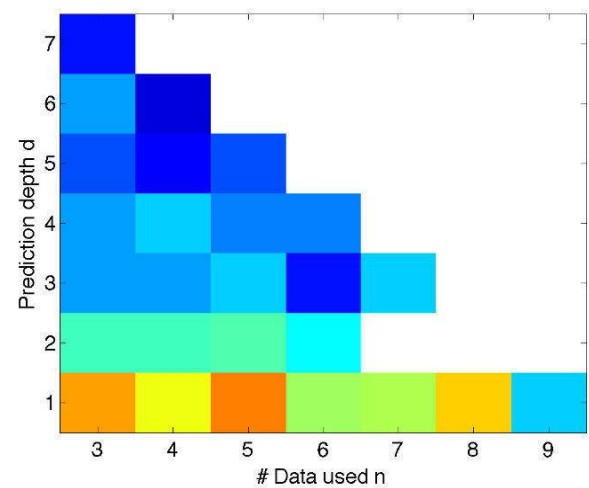

### Gompertz
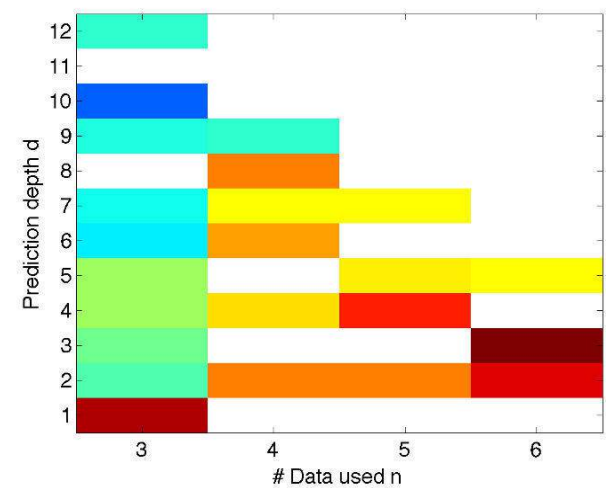

### Von Bertalanffy
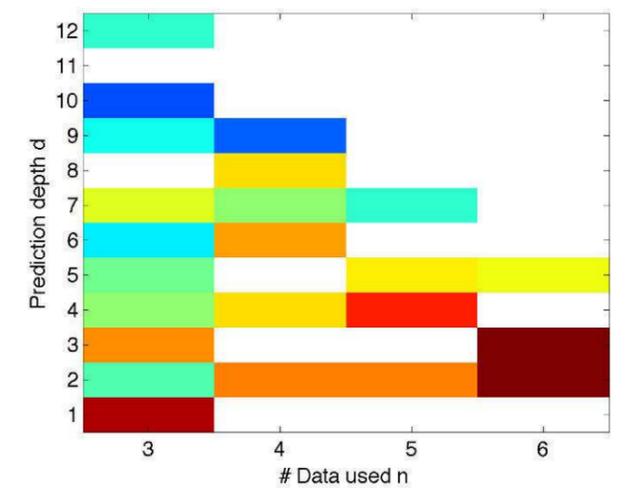

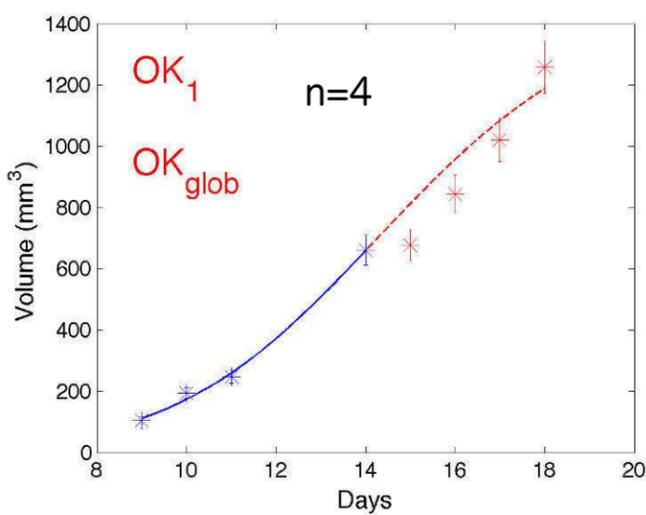

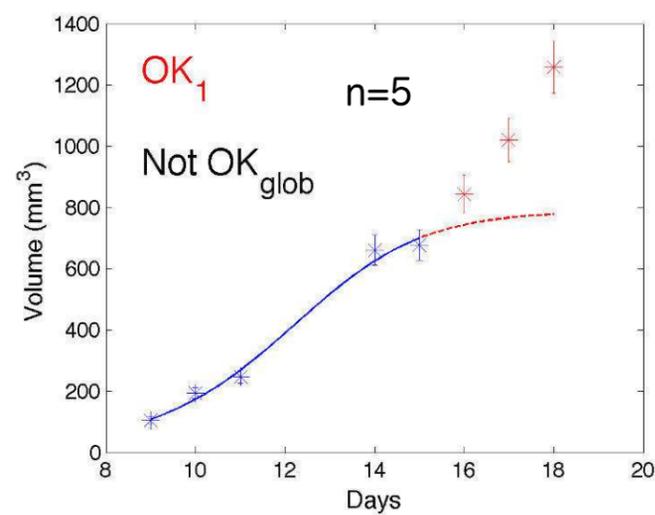

**Figure S4: Prediction.** Top: Prediction success for models that were not reported in Figure 4. Bottom: Example where prediction was less successful when using n=5 data points than when using n=4 data points, with the generalized logistic model.

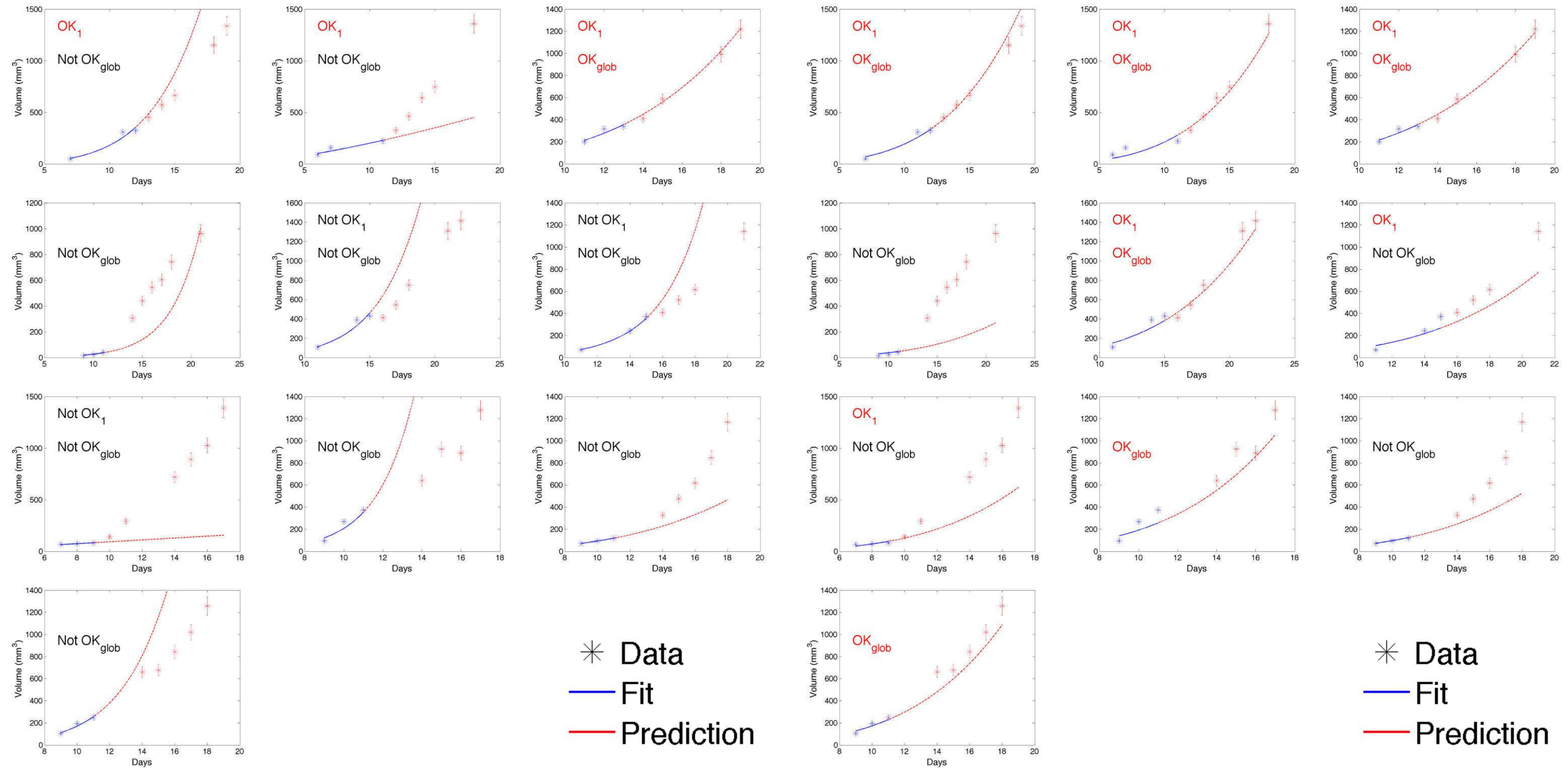

**Figure S5: Forecast improvement of the power law model when using a priori information and the lung data set.** Fits were performed using the first three data points, for each animal. A priori information (learned on a different data set) was added during the fit procedure for the predictions on the right. Shown is a particular replicate among the 100 subdivisions of the global group (20 mice) into one "learning" group and one "forecast" group.